# On Physical Origins of Learning

**Alex Ushveridze**

Algostream Consulting, Minneapolis, MN, USA
alexush@algostream.com

*Abstract:*

*The quest to comprehend the origins of intelligence raises intriguing questions about the evolution of learning abilities in natural systems. Why do living organisms possess an inherent drive to acquire knowledge of the unknown? Is this motivation solely explicable through natural selection, favoring systems capable of learning due to their increased chances of survival? Or do there exist additional, more rapid mechanisms that offer immediate rewards to systems entering the 'learning mode' in the 'right ways'? This article explores the latter possibility and endeavors to unravel the possible nature of these ways. We propose that learning may have non-biological and non-evolutionary origin. It turns out that key properties of learning can be observed, explained, and accurately reproduced within simple physical models that describe energy accumulation mechanisms in open resonant-type systems with dissipation. Usually, the term 'learning' is associated with information acquisition and processing, and its end-goal, usually referred to as 'understanding', is viewed as a state at which the system can demonstrate the ability of independently reproducing the acquired material by purely internal means. In our schema, both the learning process and its culmination – the state of understanding – emerge as natural outcomes of the energy accumulation and dissipation processes governed by simple physical laws. Throughout the paper we stay within the formalism of classical mechanics. This allows one to demonstrate that the intricate processes responsible for intelligent behavior do not necessarily require complex physics to explain them and can be modeled within simple physical systems of low dimension. We analyze some 'toy' models imitating the learning process, and, based on this analysis, propose a simple physical principle (which we call the 'energy flow maximization principle') allowing one to describe the processes of learning in ways compatible with both the inner logic of AI/ML algorithms, motivation-driven behavior of intelligent resource-seeking agents, and the abstract formalism of Newtonian dynamics.*

## 0. Introduction

The ability to learn is an inherent property of all life forms on Earth and a key ingredient of both genuine and artificial intelligence. In the living world, this ability is widely seen as the product of a long process of evolution, guided by natural selection, where organisms unable to learn eventually died out, while those that were able to learn not only survived but also passed on this trait to future generations. Although this argument is considered as basically correct [1,2], it may not be able to provide a complete picture. The fact is that natural selection can only explain the consolidation of skills that are in principle feasible, i.e., skills that not only do not contradict the laws of physics but are directly supported by them.

Otherwise, natural selection is powerless. For example, if the relationship between air density and gravity were different and unfavorable for the physical process we usually call 'flight', no random mutation could cause a living being to take off, despite the clear competitive advantage of such an ability. It is naturally to assume that similar argument can be applied to all other skills including such a fundamentally important and universal skill as an ability to learn. But then, it is natural to ask: what physics could be responsible for this ability? In other words, can we describe a certain natural proto-learning phenomenon easily explicable in the language of physics, while at the same time unambiguously interpretable as a learning process?

Why do we consider asking such questions important? Because it is gradually becomes clear that simply scaling up training data, increasing neural network complexity, and accelerating computations may not be the sole path to significant advancements in the fields of ML and AI [3-7]. There is a growing belief that a deeper understanding of the underlying physical mechanisms may also be necessary. This realization has sparked intense scientific research in an area known as PIML – Physics-Informed Machine Learning [8-15]. The core idea behind PIML is to incorporate the laws and constraints of physics into the machine learning framework, enabling models to capture and respect the underlying physics of the problem at hand. This concept has proven to be highly fruitful. However, the focus has primarily been on integrating existing physical knowledge into various ML models, rather than attempting to better understand the potential physical mechanisms of learning itself. The question of the nature of physical phenomena responsible for the emergence of intelligence has not been addressed in PIML at all and remains unclear.

It is crucial to emphasize that our focus here is not on learning in the narrow sense of this word, i.e., not on the ability of acquiring a certain specific knowledge in a certain specific field but rather on the general ability to learn. The distinction is similar to that between a program and operating system. The latter can execute any program, which makes it a sort of a 'universal program'. Similarly, the ability to learn is a universal skill that empowers a system to learn anything within a very broad range of different things, irrespective of their particular form or content. In this discussion, we will concentrate on this most universal form of learning, which, as we believe, serves as the foundation for its other more specialized forms. We shall refer to it as 'content-independent' learning and will focus on its most natural form – the so-called 'real-time learning' – attributable to the very first life forms.

The meta-algorithm of real-time learning (often referred to as 'sequential learning') appears quite simple and transparent in the language of ML: a system observes a sequence of external events unfolding in real-time and endeavors to model this sequence independently by trying to internally replicate its behavior. Sequential learning can be seen as a process of copying data in real time. The key idea is that if a system can by itself quickly and accurately reproduce the temporal profile of the data describing a certain external process, then it does likely understand that process. This mean that it (the system) can recognize hidden patterns in the external sequences of data, uncover the causal mechanisms underlying their generation, and based on that, make correct predictions about the next event – all qualities that allow living systems to be timely prepared for the future.

The above scheme may seem quite logical, if not for one additional, and, in our case, critical obstacle that turns the whole process into an absolute mystery. This obstacle is our requirement that the learning system we want to consider be an ordinary and relatively *simple* physical system described within the framework of the standard formalism of theoretical physics. This means that we want all the complex logical actions involved in the various stages of training (or sequential modeling) to be implemented as natural physical processes performed autonomously by the system itself, without any external instructions, programs, or inputs controlled by humans or other intelligent agents.

This situation leaves us with the following question: How exactly should a *simple* physical system look so that after placing it in a previously unknown external environment, it would immediately start performing all these complex actions collectively referred to as learning?

The key word here is '*simple*'. Note that the only autonomous systems capable of learning in real time in content-independent ways, are biological living organisms (with humans as the most sophisticated examples). Of course, these are all physical systems. But they, being multimolecular biological systems, are extremely complex from a physical point of view. This may give the impression that physics responsible for autonomous and content-independent learning must also be complex. However, this is not at all necessary. The fact is that the physical complexity of biological organisms is due not only to their ability to learn, but also to many other parallel processes, for example, those that are responsible for their structural stability and homeostasis. Thus, it cannot be ruled out that physics responsible just for the processes of context-independent learning, is very simple indeed and does not require the use of concepts specific to molecular dynamics and nonequilibrium thermodynamics. This is supported by the fact that if we accept that learning is a fundamental property of all living organisms (both biological and, possibly, non-biological), then most likely the hypothetic physical phenomenon responsible for its occurrence (which we above called 'proto-learning' phenomenon) must also be quite fundamental and widespread, i.e., simple enough and realizable in a wide range of conditions not requiring overly specialized or complex external factors. Otherwise, it couldn't be chosen by evolution as the starting point for countless evolutionary processes. If we take the above as a starting hypothesis, this should significantly narrow and at the same time simplify the search for the physical origins of learning.

As a first step in this search let us explicitly formulate the main conceptual difficulty we need to overcome. This difficulty is linked to the very objective of learning, as it is understood in the data science, which is the maximization of the similarity between the original external data and its copy generated by the learning system. The quantification of this similarity – building the objective function of learning – is not a big problem in ML and finding this maximum is what in ML is usually called 'training' or 'tuning'. The point is that this process is known to be the most labor-intense part of ML, as it requires making numerous iterative changes in system's internal architecture, which may require significant investment of resources and losses. But how the system may compensate for all these losses and what it stands to gain in return? Without having a regular reimbursement procedure logically and naturally connected to the process of learning, it is not quite clear what may 'motivate' the learning system (if one wants to consider it as a physical system) to enter this extremely resource-intensive learning mode? This is obviously a central question in our consideration[1].

It is intuitively clear that the motivation for learning should somehow be linked to the fact that successful learning outcome is by itself a significant resource capable of eventually compensating any possible losses. This fact may seem too obvious in relation to highly intelligent learners – those who can establish a long-term causal relationship between the limited learning efforts and potentially unlimited future payoffs. But what about the not-too-intelligent or not-yet-intelligent learning systems? How can they know in advance about future rewards at the time of just entering the learning mode? What could enforce such systems to start learning and penetrate the barrier separating them from a deserved 'good life'? What if they don't have the initial resources for that? This conundrum is referred to as the 'delayed reward problem', which is a central problem in Reinforcement Learning, also posing a significant challenge to the evolutionary argument[2].

---

[1] Of course, this is a jargon: the behavior of physical systems cannot be driven by any 'motivations'. Similarly, we cannot say that system 'wants' or 'tries to do' something. Our systems do not have any inner life and cannot have 'goals' or 'plans' for a better future. Nevertheless, using such words from time to time may help us to be a little bit less formal in describing subtle problems lying on the border between (artificial) intelligence and physics.

I do not think that this is an easy problem if we continue understanding learning as we understand it in ML (i.e., as a data modeling process). However, if we depart from this standard, then we immediately get a new window of possibilities. One of such possibilities (which we are going to discuss in this article) is to consider learning not as an initially designed skill to be developed, but as a very special example of *exaptation*. Remember that *exaptation* is a concept used in the evolutionary biology for describing situations when a trait that originally evolved for one purpose later becomes useful for a different purpose (see e.g. [17]). What if something like that happened in case of learning? What if the process which looks like learning (in the ML sense) had, in fact a different primary goal? Goal to which the delayed reward argument would simply not apply?

Surprisingly, there are good chances that the above conjecture is true. In this article we will try to bring reader's attention to the following statements about the possible features of 'proto-learning' – a hypothetic purely physical phenomenon – an entry point for the evolutionary processes resulting in all the currently known forms of learning:

- *The true purpose of proto learning is to get a resource.*
- *The reproduction of the external data by internal means is only a way/method to stimulate resource accumulation process, not an independent goal, as it may seem from the outside.*
- *The data reproduction and resource accumulation processes are inseparable in time. Any increase in reproduction accuracy is immediately rewarded with an extra resource.*

Collectively, the above bullet-points constitute what we can call the "immediate reward hypothesis", which we consider as a central statement in this note. We find this hypothesis attractive because it offers an appealing explanation for what motivates even not-too-intelligent learners to engage in learning. We will call this type of learning 'self-propelled' meaning that the learning itself can be treated as a resource-mining process and if done correctly should not require any additional external support. This also automatically leads us to the good/bad dichotomy in the context of learning: the learning path is good if it leads to gain of the resource and bad otherwise. In a philosophical sense, it can equate or at least bring closer the concepts of curiosity and hunger.

In the following sections, we will present arguments in support of the plausibility of the immediate-reward hypothesis and show how it could be realized in conceptually simple but still fundamental physical models – the models of classical mechanics. In a certain sense, the approach we are presenting here is an 'analog' version of the 'digital' model of an autonomous learning agent –the Autonomous Turing Machine (ATM) -- we proposed earlier in [17]. Our mechanical realization of the analog ATM we are going to present below may seem too naïve to some readers. While we fully understand that classical mechanics cannot provide full and final answers to all questions occurring in the context of learning, we feel that it is a good playground for understanding its true physical origins. There are the following main reasons for that. First, classical mechanics (at least its Newtonian version) is fully equivalent to the theory of dynamical systems of the most general form and thus can be used for describing all the aspects of computation including such of its sophisticated areas as AL and ML algorithms. Second, classical mechanics provides us with a simple mathematical framework allowing one to naturally introduce the notion of resource and represent its exchange between the system and its environment as a process resembling intelligent behavior of a certain autonomous system. And finally, classical mechanics lies in the foundation of all the theoretical physics which gives us the hopes for possible generalization of our mechanistic constructions to more realistic cases. We believe that best way of understanding the true origins of intelligence – one of the most complex and mysterious phenomena in the universe – is to look for these origins in the simplest possible physics, which, beyond any doubt is the classical mechanics.

---

[2] Recall a famous and still open woodpecker tongue problem: woodpecker tongue wraps around its skull which makes it hard to explain its appearance via small incremental evolutionary changes.

## 1. Learning as Resonance

To interpret learning as a resource accumulation process in quantifiable ways one first needs to formalize the notion of resource. In physics, the only natural candidate for such a role is energy – a universal resource for literally everything. Having access to energy, a system can transform it into a motion (by making change in other systems or in itself), and thus become a universal actor capable of a practically unlimited range of actions.

Note that learner $L$ being an open physical system, has to actively interact with the environment during the process of learning. This interaction reveals itself via external forces acting on the learner from the outside. The energy of the system does not have to be conserved during this interaction. Depending on the character of the latter, the external forces acting on $L$ can be either destructive or constructive, which, respectively, may lead to either the loss of the learner's energy or its accumulation. Imagine that $L$ seeks to increase its internal energy through some 'smart' interactions with its environment. Theoretically, $L$ has two polar options for achieving this: either by making appropriate changes in the environment or in itself. For big and powerful systems, both options may work. However, for smaller and weaker systems, the only viable option may be to simply adapt to external forces by making appropriate internal changes. But what kind of changes are necessary for adaptation, and what does adaptation entail?

The answer comes from physics. The simplest, and, simultaneously, smartest thing the learner can do is to use one of the most well-known physical effects – the resonance. This essence of resonance lies in the fact that when the motions of two interacting physical systems get somehow synchronized, it immediately results in a significant intensification of energy exchange between them. In such a scenario, the smaller system can greatly benefit from this interaction, as it can increase its internal energy by effectively drawing it from the larger system.

In educational literature, resonance is often discussed in the context of periodic external forces acting on harmonic (linear) oscillators, and its practical applications are typically limited to frequency-matching tasks. However, the resonance effect has a much broader nature than may seem from these specific applications. To demonstrate this, let us treat the learner $L$ as a purely mechanical system interacting with the environment through some dynamical variable $x$. Let's call it the position variable[3]. According

---

[3] By position we mean here the position of a certain part of the system directly interacting with the environment and thus playing the role of its interface. So, we can also call $x$ an 'interface' variable, to emphasize that there

to classical mechanics, the change in the system's internal energy over a period $dt$ is determined by the mechanical work performed by an external force $F$ acting on $x$ during that time. This work can be calculated by the formula:

$$dE = Fdx \qquad (1.1)$$

where $dx$ denotes the change of $x$ during the time $dt$. After dividing both sides of this equation by $dt$ and using the 'dot'-notations for time-derivatives we obtain

$$\dot{E} = F\dot{x} \qquad (1.2)$$

where $\dot{x}$ is the velocity of the system. How can $L$ increase its internal energy? From the equation (1.2), it is evident that the system's energy can only increase if the right-hand side of (1.2) is positive. As both $F$ and $\dot{x}$ are functions of time, the positivity of the right-hand side can only be achieved if the system's velocity $\dot{x}$ is synchronized with the external force $F$, i.e., both display similar behavior. This can be expressed symbolically as:

$$\dot{x} \sim F \qquad (1.3)$$

This formula, which we hereafter will call the 'resonance condition', provides small systems with a conceptually simple recipe for how to extract energy from the larger ones (e.g., from the external environment). To do that they simply need to try to reproduce the behavior of the external force $F$ in one of the system's internal variables, like $\dot{x}$ in this case.

From the above reasoning it follows that the resonance effect organically implements the key mission of learning: the reproduction of the external process by purely internal means. But now this mysterious action gets a simple physical explanation. Now it is clear why the small systems may 'want' to mimic the behavior of larger systems: Simply because such a 'parroting' is automatically accompanied by the accumulation of energy.

In an earlier discussion, we said that we want to position and study learning as a universal skill. Equations (1) – (3) can be considered as natural preconditions for this universality, as they are independent of the system and force involved. Formula (1.2) clearly indicates that the character of the external force and the form of the system are not relevant. In order to explain the resonance effect, there is no need to provide specific details about the system or the force. The only requirement is to ensure the positivity of the right-hand side of equation (1.2).

One of the nice features of the resonance model of learning is that even a tiny step towards the increase of the RHS of (1.2) (which is equivalent to a more accurate reproduction of the external input $F$ and hence to better learning outcomes), is immediately rewarded with an increased inward flow of energy. No waiting for completing the learning process is needed to be rewarded. In addition to that, the resonance learning does not impose any initial preconditions or barriers on learners to become eligible for receiving rewards. Indeed, as it is seen from the same equation (1.2), the synchronization does not need to be precise, and even approximate matching of the signs of $F$ and $\dot{x}$ may suffice to achieve some energy growth.

---

could be other, internal, non-interface variables, i.e., variables not interacting directly with system's environment, but responsible for forming its overall dynamics.

Despite all this learner-friendly environment, the learning is not effortless for the learner $L$. Taking care of the RHS of (1.2) (i.e., maximizing it or simply ensuring its positivity) is still a sole responsibility of the learner $L$. This may require some effort from $L$'s side – some changes in its internal organization may be needed, like adjustment of some of learner's internal parameters – the procedure that is usually called 'tuning'. Another typical problem the learner may face in real-life situations, is that it may not have any prior knowledge about the behavior of the external force $F$. To make $\dot{x}$ maximally close to $F$, the behavior of $F$ must first be understood and correctly predicted. This turns the synchronization given by equations (1) -- (3) into a standard machine learning (ML) problem.

To illustrate this, let's consider an idealized stock trading example. We can use $F$ to denote the change of the daily stock price, with $F > 0$ indicating an increase and $F < 0$ indicating a decrease. We can also use $x$ to represent the current amount of the stock in a trader's portfolio, with $dx$ representing the amount of stock bought ($dx > 0$) or sold ($dx < 0$) the day before. Therefore, the velocity $\dot{x} = dx/dt$ considered as a function of time represents trader's current strategy. Trader may choose from an infinite set of different strategies $\dot{x} = v[y]$ parametrized by any parameters $y$ which may directly or indirectly affect the price of the stock. Note that parameters $y$ may include both current and historical data. For a chosen strategy, the equation (1.1) shows the daily profit, while (1.2) represents the profit change rate. The recipe for gaining long-term profit is clear: one needs to synchronize the transaction pattern with the stock price by buying stock before its price goes up and selling it before it goes down. Practically this problem can be solved by selecting the best strategy from all the available strategies. For the trader this problem breaks down into two subproblems:

(i) Explicitly describe the set $y$ of strategies, i.e., select a model (in ML terminology).
(ii) Maximize the average profit accumulated during some time $\Delta t$ with respect to that set $y$ (or train the model, in ML terminology).

Trader's position is clear. But how do these two steps translate into the language of physics? The good news is that physics helps $L$ to solve the subproblem (i) automatically: there is no need for $L$ to select a special ML model, because the model is already given to $L$ for free: the equations of motion for the system $L$ can themselves play the role of such a model. In fact, any equation of motion for any physical system is in a sense a predictive model because its primary goal is to specify the next state of a system based on its current state. Indeed, if $L$ is a mechanical system, then its equations of motion must obviously be the Newtonian-type equations linking the infinitesimally near future velocity $\dot{x}$ of the system to its current velocity $\dot{x}$ and to the total current force acting on the coordinate $x$. The latter force is built up of the external force $F$ and the $x$-gradient of the potential $U(x, y)$, which, along with the interface variable $x$ may depend on other internal parameters $y$ characterizing the system $L$. This makes the solution of the dynamical equation for $\dot{x}$ an implicit function of these parameters $y$. The objective function – the integral of $F\dot{x}$ taken over some interval of time – is also becoming a function of $y$ and its maximization is a well-posed problem.

This brings system $L$ to the subproblem (ii) at which the maximization of the objective function must be performed. In contrast with the first subproblem which was trivial for $L$, the situation with the second one is not so obvious. It is intuitively clear that the only way for $L$ to achieve that goal without any external help is to treat the parameters $y$ as dynamical variables and let them evolve according to their own equations of motion. In other words, we want $L$ to be a self-tunable system capable of reaching the resonance condition on its own. Is it possible? In the following sections we will discuss this in more detail and show that the realization of self-tunability is possible even within a simple mechanical system, but it requires one more ingredient missing in the above scheme – the dissipation.

## 2. Learning and Dissipation

As we already noted, the learner $L$ is an open system, so as any open system, it must be dissipative. Omitting this fact would be incorrect. Incorporating dissipation is easy. One simply needs to rewrite the external force as follows:

$$F = f - \gamma \dot{x} \tag{2.1}$$

where $\gamma > 0$ is a dissipation coefficient and $f$ is a new version of $F$. After substituting (2.1) into the right-hand side of (1.2), we obtain the equation:

$$\dot{E} = f\dot{x} - \gamma \dot{x}^2 \tag{2.2}$$

As we see, the presence of the dissipation term in equation (2.2) does not alter the resonance condition (1.3). It merely substitutes the original external force $F$ with its corrected version $f$, but now this condition becomes more quantifiable and much more informative. By maximizing the RHS of (2.2) one can easily deduce that the maximal energy accumulation rate occurs when

$$\dot{x} = (2\gamma)^{-1} f \tag{2.3}$$

The condition (2.3) is very informative for two reasons. First, it confirms our earlier statement about the duality between the data-based and energy-based aspects of learning, by explicitly demonstrating that the condition of the maximal energy accumulation rate results in the condition of the maximal accuracy of reproduction. Second, it gives us the value of this maximum, which is

$$\dot{E} = (4\gamma)^{-1} f^2 \tag{2.4}$$

Note that this is possible only in the presence of dissipation.

Formulas (2.3) and (2.4) allow one to approach the problem of measuring the effectiveness of learning in quantifiable ways. This problem is of the highest practical importance and is closely related to question of when the self-tunability process must be stopped. How can the learner know that its learning goal is already achieved, and further improvements are impossible? We need to somehow introduce the notion of 'maximal learning effectiveness' (MLE) in the ways correctly reflecting the duality between the data- and energy aspects of the process of learning. And it turns out that without the dissipation we could not even state the problem of the maximum learning effectiveness (MLE) in the meaningful ways. Indeed, the energy-based criterion of MLE cannot be quantified in principle, because the RHS of equation (1.2) is not bounded from above. At first sight, for the data-based criterion of MLE we could say that it can be quantified as an exact match between the original data $F$ and its copy $\dot{x}$. However, this possibility is not feasible either because $F$ and $\dot{x}$ represent two different physical quantities, the force and velocity, having two different physical dimensions. This makes their direct comparison impossible. The best one can do in this case is to manually introduce a certain positive dimensional conversion coefficient between $F$ and $\dot{x}$ linked to the basic properties of the learning system and playing the role of a key parameter characterizing learning dynamics. We can call such a coefficient 'the learning constant'. However, without dissipation the introduction of such coefficient would look highly unnatural.

Note that the maximal energy growth rate given by formula (2.4) is an important milestone in the learning process. However, it is quite evident that sustained growth at this rate is impossible because any ongoing increase in $E$ would lead to an increase in the system's velocities $\dot{x}$. This, in turn, would violate the condition (2.4) at which the RHS of (2.2) is maximal, it will drop, and ultimately the growth of $E$ will slow down. This means that after reaching the point of maximal growth, the energy $E$ will continue to increase but eventually should stabilize and reach a plateau. This may happen when the entire right-hand side of (2.2) vanishes, i.e., when

$$\dot{x} = \frac{f}{\gamma} \qquad (2.5)$$

This leads us to the second significant milestone - the point at which the energy exchange between the system and its environment stops, and theoretically, the system can detach itself from the environment and continue functioning as an isolated system. In this case, we could say that the objective of learning is achieved.

As we see, both cases (2.3) and (2.5) realize the data-based MLE condition with two different conversion coefficients between $\dot{x}$ and $F$. In fact, the conversion coefficient ensuring the positivity of the RHS of (2.2) and satisfying the data-based MLE condition can be any in the range between 0 and $\gamma^{-1}$. This scale-invariance of the data-based MLE suggests that the universal data-based learning effectiveness metric applicable to all types of $\dot{x}$ and $F$ and reaching maximum at both (2.3) and (2.5) should also be scale-invariant. To define such a metric let us introduce a special bracket notation $\langle \phi \rangle$ for an average of a certain function $\phi$ of time taken over a sliding time window $\Delta t$ significantly larger than the characteristic length of the interval on which the function $f(t)$ changes its sign:

$$\langle \phi \rangle (t) = \frac{1}{\Delta t} \int_{t-\Delta t}^{t} \phi(\tau) d\tau \qquad (2.6)$$

Using this notation, we can define several basic characteristics of the learning process. This includes: the accumulated energy $A$, the inward flow of information-bearing energy $I$, the outward flow of dissipated energy $D$, the accuracy of copying $C$ and the learning time $T$.

Let us start with copying accuracy playing the role of data-based learning effectiveness. We define it as the correlator:

$$C = \frac{\langle f \dot{x} \rangle}{\langle f^2 \rangle^{\frac{1}{2}} \langle \dot{x}^2 \rangle^{\frac{1}{2}}} \qquad (2.7)$$

From the obvious inequalities it follows that this coefficient ranges between $-1$ and $+1$ and its maximum is achieved when $\dot{x} = \lambda f$ with arbitrary $\lambda > 0$. Such a metric can be useful for evaluating the progress of learning at its different stages – from very beginning to the very end.

The energy-based learning effectiveness metric is defined as the average $\langle E \rangle$ of $E$ whose time-derivative is given by the RHS of equation (2.2). To better understand how the whole learning path may look like, we can take the $\langle \ldots \rangle$-average all terms of equation (2.2). This averaging process will replace the original terms of equation (2.2) with their slower version

$$\frac{\Delta E}{\Delta t} = \langle f\dot{x}\rangle - \gamma\langle\dot{x}^2\rangle, \tag{2.8}$$

allowing us to focus on trends. Note that formula (2.8) represents the energy changing rate as a difference of two energy flows: the inward energy flow associated with the structured information-bearing forces

$$I = \langle f\dot{x}\rangle \tag{2.9}$$

and the outward energy flow associated with the unstructured dissipative forces

$$D = \gamma\langle\dot{x}^2\rangle \tag{2.10}$$

To gain a deeper understanding of the dynamics of $E$, it is useful to note that the kinetic and potential energies of closed and quasi-closed systems are typically of the same order of magnitude, as they are constantly transforming into each other as the system moves. Since kinetic energy is typically quadratic in system velocities, and the sum of kinetic and potential energies constitutes the total energy, we can use the following approximation:

$$\frac{m\langle\dot{x}^2\rangle}{2} \approx \frac{\alpha}{2} E \tag{2.11}$$

in which $m$ is the mass and $\alpha$ is a certain dimensionless coefficient depending on some internal parameters of the system. We hereafter will call it the virial coefficient because in some cases it is possible to derive precise values of α using the so-called virial theorems []. For our purposes, the approximations are sufficient, as they prioritize generality. We also assume that the data-based learning objective is 'almost' reached, so the RHS of (2.7) is positive: $C > 0$. Expressing the term $\langle f\dot{x}\rangle$ through other terms of (10), and substituting the above approximate expressions for $\langle\dot{x}^2\rangle$ and $\langle f\dot{x}\rangle$ into (2.7) we obtain the approximate equation for the averaged energy $E$:

$$\frac{\Delta E}{\Delta t} \approx m^{-\frac{1}{2}}\alpha^{\frac{1}{2}}C\langle f^2\rangle^{\frac{1}{2}}E^{\frac{1}{2}} - m^{-1}\alpha\gamma E \tag{2.12}$$

The solution of (2.12) obtained under condition that the learning starts 'from scratch', i.e., the initial value of energy $E$ at $t = 0$ is zero, reads:

$$E \approx E_{max}\left(1 - \exp\left(-\frac{t}{T_{lrn}}\right)\right)^2 \tag{2.13}$$

Here

$$T_{lrn} = \frac{2m}{\alpha\gamma} \tag{2.14}$$

and

$$E_{max} = C^2 \frac{\langle f^2\rangle}{\alpha\gamma^2} m \tag{2.15}$$

are two important constants characterizing the process of learning and both related to the dissipation coefficient $\gamma$. The first constant, $T_{lrn}$, determines the learning time – i.e., the average time needed for the system to learn the external signal $f$. It's interesting that the learning time does not depend on the character of signal $f$. As we see, this time is proportional to the mass $m$ of the system, which is quite obvious because the mass is the measure of system's inertia, and it is natural to expect that more inertial systems may need more time to learn something. But the learning time is also inverse proportional to the dissipation coefficient $\gamma$. This may seem not so obvious at first sight, but we will return to this question later. The second constant, $E_{max}$, defines the maximal energy that can be accumulated by the system during the learning process.

Based on the structure of equation (2.8), which has the form of a typical equation arising in optimal control theory, the point at which the accumulated energy reaches its maximum looks like a stable point or, more precisely, as an attractor-type dynamical equilibrium for the learning process. At this point, the two (inward and outward) flows of energy represented by the first and the second terms of the RHS of (2.8) coincide and also reach their maximum, which is given by:

$$D_{max} = I_{max} = C^2 \frac{\langle f^2 \rangle}{2\gamma} \tag{2.16}$$

We see that the system which that has finished the learning process can utilize two distinct forms of resources: (a) the static resource, represented by the accumulated energy $A_{max}$, which is immediately available for system's use, and (b) the dynamic resource, represented by the two (inward and outward) energy flow rates $I_{max}$ and $D_{max}$, which determines how quickly the static resource can be replenished in the case of overconsumption. Both resources are essential for the system's survival and effectiveness. One can draw a metaphorical parallel between these resources and personal finance, with the total accumulated energy being akin to savings and the energy flow rate being akin to income.

As seen from formula (2.15), the energy-based MLE is inversely proportional to the square of the dissipation coefficient $\gamma$. So, the smaller dissipation is, the larger is the potential energy outcome of learning. The opposite situation is with data-based learning whose effectiveness is directly proportional to $\gamma$: the larger the dissipation is -- the higher is the data replication accuracy. Formally this follows from the fact that if $\gamma \to \infty$, then the solution of the equations of motion become expandable in inverse powers of $\gamma$ and can formally be represented as $\dot{x} = \gamma^{-1} f + O(\gamma^{-2})$, which, in turn, will result in learning constant $C$ tending to 1 as $c = 1 - O(\gamma^{-1})$.

Interesting that this result is independent on the character of the learning system, so the learning appears in this case universal – any system can learn literally any pattern. Unfortunately, this fact does not seem to have any practical importance, because energy outcomes of such a universal data-based learning are negligibly small. So, we deal here with a sort of a duality which lies in the opposite and complementary role of the large and small values of $\gamma$ for the data- and energy-based effectiveness of learning. The meaning of this duality is not completely clear to us, although on a purely formal level it can be confirmed on many concrete examples.

Note also that both the static and dynamic resources are proportional to the square of the learning effectiveness coefficient $C$. This means that any improvement of the approximate equality $\gamma \dot{x} \approx f$ would lead to the increase of values of $c$ and thus lead to the further increase of both $E_{max}$ and $T_{lrn}$. In the (idealized) limiting case when $\gamma \dot{x} = f$ and, correspondingly, $C = 1$, the values of both $E_{max}$ and $T_{lrn}$ would reach their (absolute) maxima.

Of course, the latter situation is purely hypothetical (in real situations $C$ is always less than 1), but it allows us to formulate the following four maximal principles characterizing the (possibly idealized) stability point at which the balance between the outward flow of dissipated unstructured energy and the inward flow of injected structured energy occurs. At this point:

- The accuracy of the copying process $C$ is maximal,
- The outward flow of dissipated (unstructured) energy $D$ is maximal.
- The inward flow of injected (structured) energy $I$ is maximal.
- The accumulated total energy $E$ reaches its absolute maximum and stops changing.

Thus, this point represents the endpoint of learning, as no further improvement in accuracy or energy accumulation is possible. As we said above, at this point, the system essentially becomes isolated from the environment and can continue functioning on its own.

## 3. The Physics of Self-Tuning

While it may seem that the energy-based considerations given in the previous section explain a lot and bring us closer to understanding what the actual drivers of learning might be, it still remains unclear how the aforementioned processes can be triggered and then controlled in real physical systems. We are talking, of course, about the internal mechanisms that push the system towards a resonant state by making appropriate changes to the internal structure of the resonator - a process that in both ML and physics of resonators is usually called "tuning". For tuning to be possible, the resonator must be equipped with an appropriate control mechanism capable of adjusting the internal parameters of the system to their resonant values. Obviously, here we are interested only in internal control mechanisms. In other words, we want the controller to be part of the system. To achieve this, we need to find the conditions under which the system could configure itself, that is, be functionally autonomous. We know that such autonomy is quite possible (from a purely energy point of view), because some of the energy stored by the resonator of the system can be used to power its controller.  But how to implement such a scenario in practice? How to make the resonator and controller two organic parts of a single physical system?

The answer to this question depends on what we mean by a 'physical system'. If we do not impose any special restrictions on this concept and include in it any physical (i.e., not directly controlled by man) mechanisms, then there will be no problems at all.

Indeed, we can easily imagine a controller consisting of several sensors, actuators and computing devices, i.e., a fully automated device that periodically measures the level of stored energy and tries to maximize it by changing the parameters of the system in accordance with a certain optimization method. Having such a computerized controller should cost the system almost nothing, because the energy required to power it can be arbitrarily small and can be considered as an insignificant part of the dissipated energy.

Even though the above scenario may seem convincing, offering a concrete and purely physical implementation of a self-adjusting learning system, we do not feel it satisfying. The fact is that the need to use two or more different physical theories to explain one physical phenomenon may simply be a sign of our inability to fully understand the latter. As a rule, the most fundamental natural phenomena can be described simply and compactly within a single mathematical/physical structure. Moreover, sometimes such descriptions can be provided independently in several languages. For example, the idea of computation can be stated either in the language of electronics or in the language of mechanics, but we are not in a situation where we need to use a combination of both mechanics and electronics to explain calculations. In the same way, since we believe that learning can be regarded as a truly fundamental phenomenon, it is natural to expect that it must be explained within the framework of some unified and logically closed physical theory. We have already begun to use the language of mechanics, which, on the one hand, is quite simple, and on the other hand, can be considered as a prototype of many other seemingly unrelated branches of classical physics. Therefore, it will be natural if we continue to use this language of mechanics further. Moreover, it would be nice if we could limit ourselves to models with only several degrees of freedom. We are referring to some analogue of the metaphorical model of the "hydrogen atom" in physics or the application of "hello world" in programming - the simplest examples that, if properly understood, can give a clear idea of what a general case in the corresponding discipline might look like.

A possible objection to the idea of focusing on classical mechanics may be based on the argument that learning, as a generalized copying process, is one of the main forms of self-organization, which has hitherto been considered a field of macroscopic physics, describing systems with an infinite number of degrees of freedom (see, for example, [18] and references therein). Indeed, the $XX$ combination is more ordered than the $XY$ combination, so the $XY \rightarrow XX$ copying process should increase the order. According to the second law of thermodynamics, any increase in order must be accompanied by an increase in disorder. Therefore, any intelligent action (including learning) must accelerate the process of entropy production, thereby supporting the second law of thermodynamics. Some theories, such as the principle of maximum entropy production, suggest that any physical system, regardless of whether it is animate or inanimate, intelligent or unintelligent, prefers those paths of evolution along which the rate of entropy production is maximal (see, for example, [19]). But how can all these inherently multiparticle processes be effectively emulated, say, in a mechanical system with a finite number of particles?

A possible response to this objection may be as follows. Although the principle of maximum entropy production looks attractive and seems to be conceptually correct, it is still considered as a purely qualitative principle – as a certain macroscopic summary of a very complex microscopic dynamics. For that reason, it by itself cannot be considered as a fundamental driving force capable of explaining the nuances of microscopic processes. In particular, it is not immediately clear how to practically use it for constructing the objective function of learning[4], and then how to use the latter for describing the dynamics of the self-tuning process. At the same time, as we have seen above, classical mechanics, at least in its Newtonian form, seems to be quite capable of at least formulating the same questions by using the concept of friction - the microscopic realization of dissipation. The most attractive feature of friction is that being in itself a many-particle macroscopic phenomenon, it can be effectively described in purely microscopic terms, as a force acting on a single mechanical particle. The analysis given in the previous sections showed that learning and dissipation always appear together within the framework of mechanical models. We even indirectly confirmed the principle of maximum entropy production by showing that the maximum copying accuracy corresponds not only to the maximum average stored energy, but also to the maximum average energy dissipation rates $D = \gamma \langle \dot{x}^2 \rangle$.

---

[4] From a philosophical point of view, the idea of using the level of dissipation as an objective function of learning might look very attractive to us, for the reason that dissipation does not necessarily imply the direct and immediate conversion of energy into heat. This may be a more complex process that involves the deliberate conversion of the kinetic energy of the system into some other (intermediate) forms that can be used by the system for its own needs. For example, a bicycle with an electric current generator attached to its wheel may experience mechanical resistance that feels like ordinary friction, but, it is simply the effect of converting mechanical energy into electric current energy outside the formal realm of a purely mechanical system. In this sense, within the formalism of classical mechanics, dissipation can simply be used as a convenient language for describing energies whose balance cannot be effectively described within the framework of the primary mathematical formalism of the system. In this sense, if we want to allow the system to use, in whole or in part, the energy it accumulates during the learning process for any purpose other than simply maintaining the learning process, then the only channel that allows us to naturally count on such use in a mechanical model is dissipation realized through the friction forces.

The fact that the dissipation, while being an inherently macroscopic process, can effectively be described within the formalism of microscopic classical mechanics (which, in turn, seems to be fully aligned with the maximum entropy production principle in connection to the learning process), encourages us to stay within this classical formalism, hoping that it also will somehow be capable of interpreting the average dissipation rate $D$ as an objective function of learning. But how to practically approach this problem? There are two extreme and complementary architectural ways of doing that.

One way is to consider the system as a single tunable and flexible resonator endowed with a highly sophisticated controller capable of tuning the resonator for a wide range of different forces. We can call this model memoryless, because the system is not expected to physically memorize the external patterns – instead it uses the controller to simply reconfigure itself each time when a new pattern appears. If this process of reconfiguration is fast and sensitive enough – then the model could be highly effective and universal even despite its conceptual architectural simplicity. However, the price for that effectiveness might be too high because the architecture of both the resonator and controller could be too complex because of the required universality.

Another way is to consider the system as a collection of a very large number of very simple but highly specialized resonators, each pre-tuned for one type of external forces and reacting to it only. The idea is that the cumulative effect of such resonators may be considered as an energy feeding mechanism for the entire system. This way may seem attractive because it eliminates the need for controllers. The rigid configuration of resonators allows one to treat them as elementary memory units each storing a very special type of pattern capable of activating this particular resonator. Although, it may seem unclear how the required pre-tunning of resonators could be realized in practice, this problem may theoretically be overcome by completely avoiding the pre-tuning stage. The needed effect can be achieved through using the massive but fully random procedure of creating the system of such resonators. The replacement of the costly purposeful tuning process with a massive but random selection of parameters would allow one to achieve the needed functionality in relatively cheap ways. The above two ways are schematically shown in the pictures below.

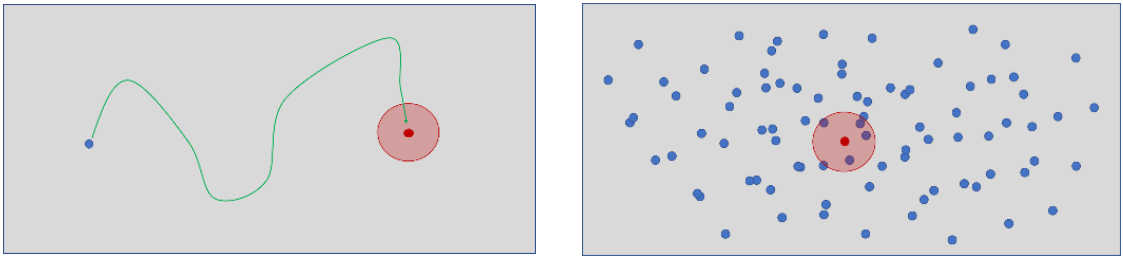

The gray areas symbolize the space of patterns to be recognized. The blue dots represent patterns the resonators can naturally react to. The red dots represent the patterns of external forces acting on the system and pink halos around them outline their neighborhoods in which the resonators can effectively fire. The first picture describes the memoryless model in which the role of the controller is crucial. The green arrow in this picture symbolizes the path in the space of all patterns along which the controller drives the resonator when bringing it closer to the external pattern. We showed this path as rather long and curvy just to stress the fact that tuning is hard – it is the most complex part of the learning process especially if the initial (waiting) pattern of the resonator and the pattern of the external force are strongly dissimilar. The second picture describes the alternative model with memory represented by a pretty dense cloud of resonators – the distance between them in the space of patterns is so small that for any external pattern there can always be found patterns that react to that pattern and fire. In that case the controllers are not needed.

Note that the two opposite scenarios described above not necessarily mutually exclusive. They may simply coexist and complement each other. The optimal learning system may contain the elements of both models. We can call it the hybrid model. For example, we may imagine a system consisting of a moderate number of simple resonator-controller pairs capable of performing a very imprecise initial signal recognition in a wide range of patterns, and then increasing the precision via some straightforward and not-too-sophisticated tuning, as illustrated in the picture below:

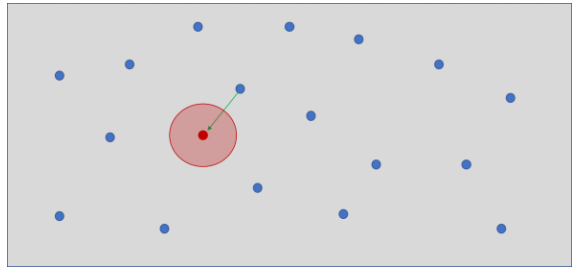

In any case, uncovering the physical mechanisms of both the memory-based and memoryless models – the two essential and complementary elements of the learning model -- would be instructive for better understanding its optimal (hybrid) architectures.  Below, in the next two sections, we consider the simplest models in which the two above scenarios can be realized in their purest form.

## 4. Attempts of Formalization and the Simplest Learning Model

To start talking of the phenomenon of learning in the language of formal models, one needs to select an appropriate mathematical framework. As such we will consider the dynamics of certain initially stationary and conservative multi-particle mechanical system perturbed by non-stationary and non-conservative forces of two types: the information-bearing external forces acting directly on particles' positions, and dissipative forces linear in particles' velocities.  Let us represent the equations of motion for our system in the following form:

$$\frac{d}{dt}\frac{\partial L}{\partial \dot{x}} - \frac{\partial L}{\partial x} = f - \Gamma \dot{x} \qquad (4.1)$$

in which $L = L(x, \dot{x})$ is a Lagrange function depending on $x$ and $\dot{x}$ – the vectors of dynamical variables and their time-derivatives, $f$ is a vector of external forces acting on the system and $\Gamma$ is a certain non-negative-definite constant matrix characterizing dissipation. We assume that this Lagrange function does not explicitly depend on time, so all the non-conservativity of this equation is concentrated in its RHS. In the absence of the latter, the energy of the system would be defined in the standard ways as

$$E = \dot{x}\frac{\partial L}{\partial \dot{x}} - L \qquad (4.2)$$

and would be a conserved quantity. However, in our case the RHS is not required to be zero, so the energy of the system is not generally conserved. Its behavior can be obtained by multiplying both sides of (4.1) by $\dot{x}$ and using (4.2). This gives

$$\dot{E} = f\dot{x} - \dot{x}\Gamma\dot{x} \qquad (4.3)$$

The above schema is general enough to allow one to formalize the problem of finding the conditions on $f$ and $L$ at which the resonance and self-tunning processes may occur in natural ways. From the energy-based standpoint, these states are characterized by the established dynamical equilibrium between the system and its environment, at which the inward and outward energy flows become balanced and the energy accumulation process stabilizes. In multi-dimensional case, this energy-based condition has a scalar nature, and therefore cannot be used for deriving the multiple force-based conditions, having the vector form. In this case, the fact of energy balance established between the system and its environment does not necessarily mean that the system is in resonance. However, the converse statement is true. Any system capable of reproducing the external forces correctly automatically becomes dynamically stabilized with the environment and its energy stops changing. This leads us to conclusion that the force-based picture is primary and the general framework we are talking about must be formulated on the level of forces, or, in other words, on the level of dynamical equations.

Taking the above into account we can formalize the notion of the resonant state $x = x[f, L]$, depending on both the character of external forces $f$ and Lagrangian $L$. We define it as a state for which both sides of equation (4.1) are simultaneously small enough:

$$\frac{d}{dt}\frac{\partial L}{\partial \dot{x}} - \frac{\partial L}{\partial x} = \epsilon, \quad f - \Gamma\dot{x} = \epsilon, \quad |\epsilon| \ll |f| \qquad (4.4)$$

Here $\phi$ is a certain limited function of time. For the learning effectiveness constant we get:

$$C = 1 - \frac{\langle f^2 \rangle \langle \epsilon^2 \rangle - \langle f\epsilon \rangle^2}{\langle f^2 \rangle^2} > 1 - \frac{\langle \epsilon^2 \rangle}{\langle f^2 \rangle} \to 1, \text{ if } \epsilon \to 0 \qquad (4.5)$$

We see that it is maximal (equal 1) only if $\epsilon = \alpha f$ or $\epsilon = 0$. For the energy stabilization condition we respectively get

$$\dot{E} = \epsilon \Gamma^{-1} f \to 0, \text{ if } \epsilon \to 0 \qquad (4.6)$$

Note that according to the above formal definition the 'null state' (i.e., state with no motion $\dot{x} = 0$ and no force $f = 0$) can also be considered as a particular case of the resonant state (the fact that in that case $C = 1$ shouldn't confuse us if we treat the statements 'system does not know anything' and 'system knows nothing' equivalent). In this context the learning process can be defined as a transition from the null to a non-null resonant state. This automatically leads us to the notions of forgetting, the transition from a non-null to the null resonant state, and re-learning, the transition between the two non-null resonant states. While learning is associated with the accumulation of energy, forgetting is an opposite process when the system loses all the accumulated energy. As to the procedure of re-learning, it may lead to both the increase and decrease of energy, depending on the character of forces in the beginning and at the end of the process. Learning, forgetting and re-learning are the three conceptually inseparable processes which must be considered together within any more or less general formalism.

Actually, the phenomenon we want to study here is the transition between different resonant states driven by changes occurring in patterns of external forces. The key word here is the adjective 'resonant' because without this requirement (assuming the smallness of $\epsilon$), the above transition process would look trivial. Indeed, any change in forces $f: f_1 \to f_2$ would automatically trigger the change in trajectories $x: x[f_1, L] \to x[f_2, L]$, and, in turn, in functions $\epsilon: \epsilon[f_1, L] \to \epsilon[f_2, L]$. In this sense, any change on the external force $f$ can be considered as a driver of changes in dynamics of $L$, by its very nature. However, the whole point is that we want $f_1$ and $f_2$ to be resonant forces, with small values of functions $\epsilon[f_1, L]$ and $\epsilon[f_2, L]$. This is what makes the problem meaningful, because this is what guarantees the high quality of learning, high values of accumulated energy and large dissipation rates. And this is also what makes the problem non-trivial because this smallness is not something that can automatically be guaranteed for any external force $f$ and for any Lagrangian $L$.

This leads us to the following two problems for a given a set of external forces $f = \{f_1, \dots, f_N\}$:

1. What is the minimal Lagrange function $L$ for which all functions from the above set can induce the resonant states $x = x[f, L]$ with sufficiently small $\epsilon = \epsilon[f, L]$?

2. What is the minimal Lagrange function $L$ for which the transitions $f_i \to f_k$ can induce transitions $x[f_i, L] \to x[f_k, L]$ with sufficiently small $\epsilon[f_i, L]$ and $\epsilon[f_k, L]$?.

Note that these two problem are relevant to both learning, forgetting and re-learning. This formulation seems pretty general and, as we see, does not even require the explicit distinction between the resonator and controller.

Among the many possible examples of resonance states, we want to single out one for which the function $\epsilon$ degenerates to an identical zero, $\epsilon = 0$, when both RHS and LHS of equation (4.1) disappear at the same time. This is indeed a very interesting case, because in it we are dealing with the behavior described by the standard Lagrange equations (with zero RHS). This makes it indistinguishable from the behavior of a completely isolated and conservative system. Next, we will refer to states with $\epsilon = 0$ as "pure resonance states".

In a sense, the situation with pure resonance states resembles the situation with the eigenvectors of matrices: there are some vectors on which the action of the matrix is trivialized and reduced to a simple multiplication by a number. Here we have a similar situation. There are trajectories for which the behavior of an initially open and non-stationary system is trivialized, and the latter begins to behave as closed and stationary.

Although pure resonance states are a kind of idealization, they are useful for developing a variety of toy learning models. In this connection, it would be natural to ask about the simplest systems that illustrate the most fundamental characteristics of learning in terms of pure resonance states. We are referring to a kind of idealized models of the "elementary student", which could look simple, but at the same time be quite rich and informative.

Note that the complexity of the system characterized by the Lagrange function $L$ is dictated by the complexity of the information-bearing external function $f$ acting on the system. The more intricate the behavior of $f$ becomes, the greater the challenge in comprehending and learning it, necessitating the use of more sophisticated Lagrangians for this purpose. It is naturally to expect that the simplest learning model must be associated with a simplest external force – a constant

$$f = const, \tag{4.7}$$

whose value, $\rho$, is the only parameter that needs to be learned. Indeed, it is easy to see that the role of $L$ in that case can be played by the Lagrangian of a free particle

$$L = \frac{m\dot{x}^2}{2}, \tag{4.8}$$

characterized by a single parameter $m$ (the mass). The damped version of the equation of motion for such a system, perturbed by a certain constant external force, has the form:

$$m\ddot{x} = f - \gamma\dot{x} \tag{4.9}$$

The constant force is the simplest data that can be learned by the system with 100% accuracy. Indeed, in case when force is constant, any function nullifying the RHS of (4.9) nullifies its LHS too. Thus, it represents the pure resonant state that we talked about earlier. The key characteristics of this state including accumulated energy $E$, dissipation rate $D$ and learning/copying effectiveness constant $C$, respectively, are given by

$$E_{max} = \frac{mf^2}{2\gamma^2}, \qquad I_{max} = D_{max} = \frac{f^2}{\gamma}, \qquad C = 1 \tag{4.10}$$

Now assume that the initial external constant force $f = f_1$ suddenly changes at time $t = 0$ to another constant force $f = f_2$. This process can be described by a time-dependent function

$$f(t) = f_1 + (f_2 - f_1)\theta(t) \tag{4.11}$$

where $\theta(t)$ denotes the so-called theta-function, which is 0 for $t < 0$ and 1 for $t \geq 0$. Since the equation (4.9) is explicitly solvable for any function $f$, we can simply write down its solution for (4.11):

$$\dot{x} = \frac{f_1}{\gamma} + \frac{f_2 - f_1}{\gamma}\left(1 - e^{-\gamma t/m}\right)\theta(t) \tag{4.12}$$

This solution describes both learning ($f_1 = 0$, $f_2 \neq 0$), forgetting ($f_1 \neq 0$, $f_2 = 0$), and relearning ($f_1 \neq 0$, $f_2 \neq 0$). The characteristic execution time for either of these processes is $T \approx m/\gamma$. The fact that all these processes are described by the same formula shows that there is no conceptual difference between them: learning can be viewed as a simple oblivion of the previous state.

## 5. Pure Resonant States and the Inverse Problem

Note that the correspondence $L \leftrightarrow \{f_1, ..., f_N\}$ between the Lagrange function and the set of forces reducing an open system to the seemingly closed form on a certain set of trajectories $\{x_1, ..., x_N\}$ can be read in both directions: as a direct problem or as an inverse problem. The direct problem is trivial. Take a Lagrangian for a certain stationary system and select a set of its trajectories $\{x_1, \ldots, x_N\}$ corresponding to different initial conditions. Find their time-derivatives, multiply them by $\gamma$ and you will get the desired set of forces $\{f_1, \ldots, f_N\}$ nullifying the RHS of the corresponding equations of motion.

The inverse problem is (as usually) much more complex than the direct one. However, it should be our primary focus in the context of learning, because the primary question related to any learning system is how it could accommodate to a given external force rather than what kind of forces it would prefer to deal with. Below we consider different types of function $f(t)$ and explore some of the possible ways of approaching this inverse problem. Our goal is to just try to understand how difficult this problem might be in principle. Our natural desire is of course to find the easiest ways which may lead to the maximally simple solutions.

## Harmonic Case

The simplest non-constant case for which the solution of the inverse problem is explicitly known is the case of harmonic external force

$$f(t) = \rho \sin \omega t \tag{5.1}$$

The role of the learner in this case is played by a one-dimensional model of the harmonic oscillator described by the Lagrange function

$$L = \frac{\dot{x}^2}{2} - \frac{\omega^2 x^2}{2} \tag{5.2}$$

If the relation

$$\dot{x} = \frac{\rho \sin \omega t}{\gamma} \tag{5.3}$$

holds, then both sides of the equation

$$\ddot{x} + \omega^2 x = \rho \sin \omega t - \gamma \dot{x} \tag{5.4}$$

vanish, which means that we deal with a pure resonant state on which the equation for $x$ reduces to the conservative form of the simple undamped and undriven (isolated) harmonic oscillator. In this case, the learning effectiveness is obviously maximal, $C = 1$, and the energy-related quantities coincide with the general formulas (2.15) and (2.16) with $\alpha = 1$ taken for the virial constant.

$$E_{max} = \frac{m \langle f^2 \rangle}{\gamma^2}, \qquad I_{max} = D_{max} = \frac{\langle f^2 \rangle}{\gamma}, \quad \text{where} \quad \langle f^2 \rangle = \frac{\rho^2}{2} \tag{5.5}$$

We see that this case is somewhat similar to the trivial case considered in previous section. Note also that this solution is stable. Indeed, if we represent $x$ as $x = x_f + \delta x$, where $x_f$ satisfies (36) and $\delta x$ is a deviation, then for the latter we obtain the equation

$$\delta \ddot{x} + \omega^2 \delta x = -\gamma \delta \dot{x} \tag{5.6}$$

whose solutions vanish as $t \to \infty$, no matter with which initial conditions they start. This means that we can treat the pure resonant state $x_f$ as an attractor.

## Periodic Anharmonic Case

Assume that the function $f(t)$ is strictly periodic (with period $2T$) and anti-symmetric $f(t_n - t) + f(t_n + t) = 0$ with respect to its zero points at $t_n = t_0 + nT$. Also allow $f(t)$ to have any behavior between these points other than changing its sign. A typical example of such function is shown below

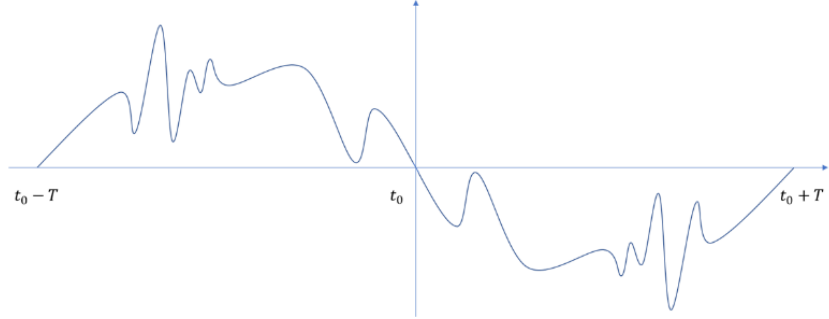

We will call such functions 'single-wave periodic functions', or simply 'single-wave functions'. We claim that that this function can be a pure resonant state for a certain Lagrange functions of the form:

$$L = \frac{\dot{x}^2}{2} - U(x) \tag{5.7}$$

The proof of the above statement is straightforward and is based on the explicit construction of the potential $U(x)$. Let us start with the equation (4.1), which in case of (5.7) reads

$$\ddot{x} + U'(x) = f - \gamma \dot{x} \tag{5.8}$$

If the maximal learning effectiveness condition (2.2) is satisfied, then the RHS of (5.8) must be zero. By integrating the equation (2.2), we can find the coordinate $x$ as a function of time $t$

$$x = x(t) = \frac{1}{\gamma} \int^t f(t) dt \tag{5.9}$$

If we want this function to satisfy the whole equation (5.8), we need to make sure that its LHS is zero too:

$$\ddot{x} + U'(x) = 0 \tag{5.10}$$

But (5.10) is nothing but the equation of motion for the stationary system with Lagrangian (5.7). This means that function

$$E = \frac{\dot{x}^2}{2} + U(x) \tag{5.11}$$

representing the energy for (5.7) must be a constant for $x = x(t)$. This information is sufficient for reconstructing the shape of the potential $U(x)$. Indeed, using (5.11) and assuming (for definiteness) that the minimum of the potential is 0, we can immediately represent the potential

$$U(x) = \frac{f_{max}^2}{2\gamma^2} - \frac{f^2(t)}{2\gamma^2} \tag{5.12}$$

as a function of $t$. To rewrite it as a function of coordinate $x$, we need to replace $t$ in (5.12) with $t(x)$ -- the inverse function of $x(t)$. Note however that $t(x)$ can be uniquely defined only on those intervals on which the direct function $x(t)$ is monotonic. But we have only two such intervals for each period: like $[t_0, t_0 + T]$ and like $[t_0, t_0 - T]$. The function $x(t)$ is monotonically increasing in one of them, and monotonically decreasing in another one. This gives us two different branches of function $t(x)$. Because of the anti-symmetry of $f(t)$, these two branches differ only by sign and thus leave the final expression

$$U(x) = \frac{f_{max}^2}{2\gamma^2} - \frac{f^2(t(x))}{2\gamma^2}, \quad x \in [-x_{max}, +x_{max}] \tag{5.13}$$

branch-independent and left-right symmetric with respect to $x$. This completes the proof.

What about the stability of the above solution $x(t)$? We can approach this question in the same way as in the harmonic case. Denoting the resonant solution by $x_f$, and considering small deviations from it, $x = x_f + \delta x$, we can write the corresponding equation for them:

$$\ddot{\delta x} + U''(x_f)\delta x = -\gamma \dot{\delta x} \tag{5.14}$$

Since $U(x)$ is an oscillatory type potential describing a finite and periodic motion in an infinite well, its second derivative $U''(x)$ must be positive (at least on average):

$$\overline{U''(x)} = \frac{1}{2x_{max}} \int_{-x_{max}}^{+x_{max}} U''(x)dx = \frac{U'(x_{max})}{x_{max}} > 0 \tag{5.15}$$

This makes the equation (5.14) similar to the equation of motion for the damped harmonic oscillator with effective eigenfrequency $\omega^2 \approx \overline{U''(x)}$ and friction $\gamma$. The amplitude of solutions $y$ of such equation should exponentially vanish which means that the resonant solution $x(t)$ we just constructed for the potential $U(x)$ is asymptotically stable.

So far, we considered a strictly periodic single-wave function of an arbitrary shape. We can claim that for any given external force $f(t)$ and any dissipation coefficient $\gamma$ it is always possible to find a one-dimensional Lagrangian $L$ for which the dynamical equation (4.1) allows solutions satisfying the relation (5.9) for a certain macroscopic time interval. However, the Lagrange function ensuring such a correspondence does not necessarily have the form (5.7). We have a general proof of this statement, however it is too technical and for that reason we will place it in Appendix A.

## Quasi-periodic forces

If we drop the requirement of one-dimensionality, the spectrum of possibilities for solving the inverse problem tremendously widens. Consider a more general case when function $f(t)$ is a superposition of a finite number of single-wave periodic functions $f_\alpha(t)$, $\alpha = 0, 1, \ldots, N$:

$$f(t) = \sum_{\alpha=0}^{N} f_\alpha(t) \tag{5.16}$$

We will allow functions $f_\alpha(t)$ to have different half-periods $T_\alpha$, different reference points $t_{0\alpha}$ and different shapes between the corresponding zero points $t_{n\alpha} = t_{0\alpha} + nT_\alpha$. Hereafter we will call the resulting function $f(t)$ the 'multi-wave quasi-periodic function', or simply 'multi-wave function'.

Can we solve the inverse Lagrange problem for such functions? The answer is yes, but if we want to continue restricting ourselves to the Lagrange functions quadratic in system velocities, we will need to extend the number of degrees of freedom in the system and consider multi-dimensional versions of equations (4.1). For making our further notations more compact let us denote the whole set of dynamical variables (coordinates) by $x_\alpha$, $\alpha = 0,1,\ldots,N$ and take $x_0$ as the primary variable directly interacting with the external force $f(t)$. Then the system of equations (4.1) can be rewritten in the form:

$$\frac{d}{dt}\frac{\partial L}{\partial \dot{x}_\alpha} - \frac{\partial L}{\partial x_\alpha} = \delta_{0\alpha}(f - \gamma \dot{x}_0), \qquad \alpha = 0,1,\ldots,N \tag{5.17}$$

where

$$L = \frac{1}{2}\sum_{\alpha=0}^{N} \dot{x}_\alpha^2 + U(x), \quad x = \{x_0, \ldots, x_N\} \tag{5.18}$$

The explicit form of (5.17) reads

$$\ddot{x}_\alpha + \frac{\partial U(x)}{\partial x_\alpha} = \delta_{0\alpha}(f - \gamma \dot{x}_0), \qquad \alpha = 0,1,\ldots,N \tag{5.19}$$

Consider the following orthogonal coordinate transform:

$$x_\alpha = \sum_{\beta=0}^{N} M_{\alpha\beta} y_\beta, \quad \frac{\partial}{\partial x_\alpha} = \sum_{\beta=0}^{N} M_{\alpha\beta} \frac{\partial}{\partial y_\beta} \tag{5.20}$$

Substituting (5.20) into (5.19) and introducing new notation

$$\gamma_\alpha = \gamma M_{0\alpha} \tag{5.21}$$

we obtain

$$\sum_{\beta=0}^{N} M_{\alpha\beta} \ddot{y}_\beta + \sum_{\beta=0}^{N} M_{\alpha\beta} \frac{\partial U(x)}{\partial y_\beta} = \delta_{0\alpha} \sum_{\beta=0}^{N} (f_\beta - \gamma_\beta \dot{y}_\beta). \tag{5.22}$$

After some transform, we get

$$\ddot{y}_\alpha + \frac{\partial U(x)}{\partial y_\alpha} = M_{0\alpha} \sum_{\beta=0}^{N} (f_\beta - \gamma_\beta \dot{y}_\beta), \qquad \alpha = 0,1,\ldots,N. \tag{5.23}$$

The potential $U(x)$ in this formula must be considered as a function of the new variables $y$. Remember that our goal is to find the form of this potential. It looks like it is easier to do that first in variables $y$ and then transform the found form to the original variables $x$. We use the following ansatz for $U(x)$:

$$U(x) = \sum_{\beta=0}^{N} V_\beta(y_\beta) \tag{5.24}$$

Substitution of (5.24) into (5.23) gives:

$$\ddot{y}_\alpha + \frac{\partial V_\alpha(y_\alpha)}{\partial y_\alpha} = M_{0\alpha} \sum_{\beta=0}^{N} (f_\beta - \gamma_\beta \dot{y}_\beta), \qquad \alpha = 0,1,\ldots,N \tag{5.25}$$

Note also that in terms of new variables the maximal learning effectiveness condition reads:

$$f - \gamma \dot{x}_0 = \sum_{\beta=0}^{N} (f_\beta - \gamma_\beta \dot{y}_\beta) = 0 \tag{5.26}$$

Now we have everything to repeat the logic used in the previous case. If we want (5.26) to be satisfied, we should require that

$$\dot{y}_\alpha = \frac{f_\alpha}{\gamma_\alpha}, \qquad \alpha = 0,1,\ldots,N \tag{5.27}$$

These conditions look as $N+1$ separate equations (41). Now the RHS of (5.25) is 0. If we want $y_\alpha$ to satisfy the whole equation (5.25), its LHS must be zero on (5.27). But this leads us to the system of $N+1$ distinct homogeneous equations

$$\ddot{y}_\alpha + \frac{\partial V_\alpha(y_\alpha)}{\partial y_\alpha} = 0, \qquad \alpha = 0,1,\ldots,N \tag{5.28}$$

which can be analyzed separately exactly in the same ways as it was done in the previous subsection. Essentially we need to reconstruct the form of the potentials $V_\alpha(y_\alpha)$ each one as a function of the transformed coordinate $y_\alpha$ provided that the form of this coordinate as a function of time is given. And we know that such a solution exists because each coordinate is associated with the single-wave function $f_\alpha$. This completes the proof.

## Systems with multiple pure resonant states

So far, we considered the cases of systems having a single pure resonant state (PRS). Now we consider systems allowing multiple PRS. The difference between these two systems is that the former may have maximum one resonant attractor (in case when the corresponding PRS is stable), while the latter may allow multiple attractors and thus learn and relearn between multiple types of data.

As we show below, building such systems is not difficult and the role of building block can be played by the single-PRS systems.

Consider $N$ Lagrange functions $L_n(x, \dot{x})$ each having a single PRS $x = x_n$ for the external forces $f_n$ where $n = 1, \ldots, N$. This means that

$$\frac{d}{dt}\frac{\partial L_n(x_n, \dot{x}_n)}{\partial \dot{x}_n} - \frac{\partial L_n(x_n, \dot{x}_n)}{\partial x_n} = f_n - \gamma \dot{x}_n, \quad n = 1,\ldots,N \tag{5.29}$$

Having these $N$ conditions, we can construct a new Lagrangian for which each of these forces will induce a separate PRS. This will give us a multi-PRS learning system. We will show that it is easy to do it by increasing the dimensionality of the system, but not significantly – adding just one variable will suffice. The resulting Lagrange function can be written down explicitly:

$$L(x, \dot{x}, y, \dot{y}) = \dot{y}^2/2 + \sum_{n=1}^{N} A_n(y) L_n(x, \dot{x}) \tag{5.30}$$

Here $y$ denotes a new dynamical variable, and $A_n(y)$ are some functions of these variables which we will specify latter. The equations of motion for this Lagrangian read:

$$\dot{y} \sum_{n=1}^{N} A'_n(y) L_n(x, \dot{x}) + \sum_{n=1}^{N} A_n(y) \left( \frac{d}{dt} \frac{\partial L_n}{\partial \dot{x}} - \frac{\partial L_n}{\partial x} \right) = 0 \tag{5.31}$$

$$\ddot{y} + \sum_{n=1}^{N} A'_n(y) L_n(x, \dot{x}) = 0 \tag{5.32}$$

Let us select $N$ values $y_k$, $k = 1, …, N$ such that

$$A'_n(y_k) = 0 \tag{5.33}$$

and

$$A_n(y_k) = A_n \delta_{nk} \tag{5.34}$$

Taking subsequently

$$y = y_k, \quad x = x_k, \quad k = 1, …, N \tag{5.35}$$

we obtain zeros in both-hand sides of both equations (5.31) and (5.32). This completes the construction of the multi-PRS Lagrangian for a given set of external forces.

The fact that a certain system allows multiple PRS does not alone guarantee that the transition between these states can be triggered only by switching between the patterns of the corresponding external forces. We had such a situation in case of the simplest (free-particle) learning model considered in section 5. This model demonstrated the ability of relearning driven by any replacement of any previously learned constant force with other constant force. Will the model (5.30) be able to relearn in similar ways? We currently do not know. With this Lagrangian we can consider the problem of finding the dynamical solutions for the learning process when the system is initially in the state with $y = y_k$, $x = x_k$, $f = f_k$, where $k = k_1$, and then the force switches to $f = f_k$ with $k = k_2$. Under which additional conditions imposed on the Lagrangians $L_n$ and functions $A_n$, the solution of this equation will asymptotically tend to $y = y_k$, $x = x_k$, where $k = k_2$? We hope to address this question in one of our further publications.

# 6. Self-Tunability: The Simplest 1-Dimensional Case

In the previous section we briefly discussed learning model capable of learning the simplest non-constant and strictly periodic external force (5.1) with shape fully characterized by two parameters, $\rho$ (the amplitude) and $\omega$ (the frequency). However, what we discussed in that section could be called the 'partial learning', as during this learning the system learned only one of the two parameters – the amplitude – and didn't change its own internal structure. The ultimate goal-maximum of learning the force (5.1) is to learn both parameters. To do that, we will start with practically the same Lagrange function as we had before in (5.2) but now having a little bit more detailed form

$$L = \frac{m\dot{x}^2}{2} - \frac{kx^2}{2} \tag{6.1}$$

with two constant parameters $m$ (the mass) and $k$ (the stiffness). The corresponding equation (36) will now take the form

$$m\ddot{x} + kx = \rho \sin \omega t - \gamma \dot{x} \tag{6.2}$$

The case of partial learning we considered earlier was characterized by condition

$$k/m = \omega^2 \tag{6.3}$$

Now let us focus on a more complex full learning case in which both the parameters $\rho$ and $\omega$ characterizing the external force $f$ have to be learned. So, we will be interested in the situation when the conditions (6.3) are not met, and the system is forced to tune in to achieve the resonant state. The only way to realize such a scenario is to allow the constants in (6.1) to change, and for that we need to treat them as dynamical variables. In our case we have two such constants, $m$ and $k$. It is important to note that theoretically their 'dynamization' can be achieved even without making any changes to the Lagrangian. We can call such a scenario the 'minimalist' approach. Its beauty lies in the fact that to make a certain variable dynamic it is sufficient to just declare it a dynamic variable, which, in practice, can be done simply by extending the original equations of motion with an additional equation of motion for the chosen variable (the former constant). Such a declaration will obviously change the equations of motion for the original variable. For example, the equation of motion (6.1) will acquire an extra term proportional to the time-derivative of $m$:

$$m\ddot{x} + kx = \rho \sin \omega t - (\gamma + \dot{m})\dot{x} \tag{6.4}$$

Note the absence of the time-derivatives of the second constant $k$ in (6.4). This follows from the general structure of the Lagrange equations for $x$, and suggests that the dynamization of only one of the two constants may suffice. For that reason, hereafter we will consider only $m$ as a dynamical variable and continue treating $k$ as a constant. To ensure that the dynamization of $m$ does not change too much the form of the original solution, we need to find conditions at which the changes of $m$ are slow. In addition to that, we need to find conditions under which the system can perform all the tuning operations by itself.

If we would wish to keep the things at maximal level of generality, we could try to extend the initial Lagrangian (6.1) with some additional terms

$$L = \frac{m\dot{x}^2}{2} - \frac{kx^2}{2} + \Delta L(\dot{m}, m) \tag{6.5}$$

including, for example, the kinetic term for $m$ of the form $K_m = \mu \dot{m}^2/2$ and maybe also some potential term $U_m = V(m)$. The inclusion of kinetic term with large $\mu$ – the parameter playing the role of $m$'s inertia – would suffice for slowing down the motion of $m$, and, respectively, minimizing the impact of the extra term $\dot{m}$ in (6.4). Noting that the parameter $m$ can also be affected by external forces and dissipation, we can write down the most general form of the dynamical equation for $m$:

$$\frac{d}{dt}\frac{\partial \Delta L}{\partial \dot{m}} - \frac{\partial \Delta L}{\partial m} - \frac{\dot{x}^2}{2} = -f_m - \gamma_m \dot{m} \tag{6.6}$$

The introduction of the dissipative terms in the RHS of (6.6) is necessary because there should be some forces stopping the motion of $m$ once the resonant state is attained. Also note that by requiring that the dissipation coefficient $\gamma_m$ is large we can slow down the motion of $m$ even without increasing the values of its inertia $\mu$.

Note that the expression for energy, will obviously acquire an additional term

$$E = \dot{x}\frac{\partial L}{\partial \dot{x}} + \dot{m}\frac{\partial L}{\partial \dot{m}} - L \tag{6.7}$$

and the formula for its time-derivative will take the form:

$$\dot{E} = \rho \sin \omega t \cdot \dot{x} - \gamma \dot{x}^2 - f_m \dot{m} - \gamma_m \dot{m}^2 \tag{6.8}$$

From this formula it follows that once the motion of $m$ stops, the last two terms in (6.8) immediately vanish and the entire expression (6.8) degenerates into its original form (2.2) we started with. The only problem with this picture is that we still need to find an appropriate extension $\Delta L$ for which such a behavior would be realizable in natural ways.

Fortunately, the solution is simpler than it may seem. It turns out that the desired behavior can be realized within the minimalist approach, at which the extra term is simply zero, $\Delta L = 0$, so the original Lagrangian (5.2) does not change. In that case $m$ becomes a 'cyclic variable', i.e., a variable without a kinetic term. In that case the dynamics of $m$ takes especially simple form:

$$\gamma_m \dot{m} = -f_m + \frac{\dot{x}^2 [m]}{2} \tag{6.9}$$

and, as we see, is governed by the dissipation coefficient of $m$. If latter is large, then the motion of the variable $m$ will be slow. The beauty of the above equation lies in the fact that it can be integrated explicitly:

$$m = m_0 - \frac{f_m}{\gamma_m}t + \frac{1}{2\gamma_m}\int_0^t \dot{x}^2 \, dt \tag{6.10}$$

Assume that learning starts at $t = 0$ from the 'null' state at which $x = 0$ and $\dot{x} = 0$. At this state $E = 0$ too irrespective of the values of constants $m$ and $k$. Assume that the initial value of $m$, which we denote by $m_0$, is large enough, larger than the resonant value of $m$ at given $k$: $m_0 > m_r = k/\omega^2$. At the initial stages when $t > 0$ but still small, the value of $m$ will drop almost linearly because $\dot{x}$ will be small and the contribution of the third term in (6.10) will be negligibly small. This motion of $m$ towards its resonant value $m_r$ will gradually slow down as the resonance effects become more apparent, and the role of the third term will increase. This motion will stop only when the growth rate of the third term in (6.10) becomes comparable with the linear drop rate of the second one. To show how this mechanism may work, let us integrate the equations over a sliding time window of the width $\Delta t$ (much larger than the period of oscillations of $x$). Using the formula (2.10) expressing the average $\langle \dot{x}^2 \rangle$ via the dissipation rate $D$, we obtain

$$\gamma_m \frac{\Delta m}{\Delta t} = -f_m + \frac{D(m)}{2\gamma} \tag{6.11}$$

We can see that (6.11) takes the form of the optimal control equation. Indeed, the dissipation term $D(m)$ depends on $m$ and reaches its maximum at the resonant state when the condition () holds. Assume that

$$f_m = \frac{1}{2\gamma} \max_m D(m) - \epsilon \tag{6.12}$$

where $\epsilon$ is a small positive constant. In that case there are some values of $m$ at which the RHS of equation (6.12) vanishes. These values will be the stable points for these equations. The best way of reaching the stable point of equation for $m$ is to start iterating it from large values of $m$ and gradually lowering them. Note the natural appearance of the maximum of the dissipation term in the above expression – one more indicator that the macroscopic quantity used in the maximum entropy production principle as objective function of learning can naturally be obtained within a purely microscopic approach.

One may ask how the mass of a mechanical body can itself be considered as a dynamical variable. The answer is simple: instead of a real mass one can consider the effective mass. The following example (which mathematically is more complex than the idealized model we considered above but may look more natural from the physical standpoint) clearly illustrates how the idea of self-tuning can be realized within a real system with variable effective mass. As such a system we can consider a pendulum with given weight of a constant mass $m$ but variable length $L$.

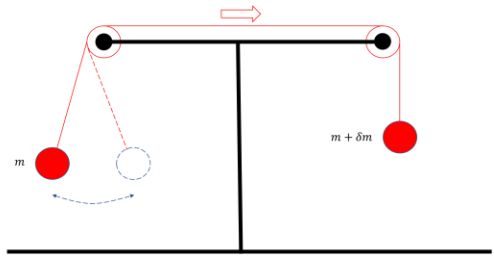

As seen from the picture, the length of pendulum is controlled by a slightly heavier counterweight of the mass $m + \delta m$, causing it to slowly move down following the gravitational forces compensated by friction and thus gradually shorten the pendulum length. If this pendulum is affected by a periodic force of frequency $\omega$, then near its resonant length $L = g/\omega^2$ the amplitude of swings will increase and due the increased centrifugal forces the effective weight of the pendulum will become equal to the counterweight and the further shortening of the length will stop. The system will become 'almost' tuned for the frequency $\omega$.

Let us return to the equation (6.12) which is pretty easy to analyze. We see that the smaller the value $\epsilon$ is, the closer can controller bring the system to the resonant state and the higher is the quality of learning that can be attained. However, the higher the quality of learning is, the harder is to maintain it. The point is that if, because of some unexpected fluctuation, the value of $\epsilon$ drops below zero, then the stable point immediately disappears and the system will enter the unstable mode: the values of $m$ will continue dropping and may even become negative, which is obviously a nonsense. The possible scenarios of such behavior are illustrated in the picture below giving a graphical version of the process of learning described by formula (6.12):

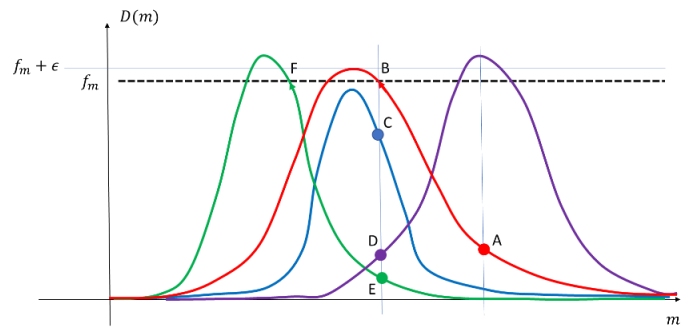

The learning starts at point A and, if the shape of the function $D(m)$ (red colored bell-shaped curve) does not change, eventually reaches the stable point at its intersection of this curve with the horizontal line symbolizing the constant external force $f_m$. However, if the frequency or the amplitude of the information-bearing external force $f(t)$ suddenly changes, it will immediately cause the changes in the shape of the corresponding dissipation curve $D(m)$. The new possible shapes are depicted in blue, purple, and green. As seen from the picture, the blue and purple cases lead to the uncontrollable move of $m$ towards the negative values. But in the green case this motion stops at the new stable point. In that case we can say that the system re-learns.

Summarizing, we can see that although the above model does demonstrate a possible scenario of self-tuning, it has an essential drawback allowing one to do that only once. Re-learning is possible, but only in some special cases when the new 'learning material' lies on the path initially chosen by the 'learner'. In the language of the variable length pendulum model, this corresponds to the impossibility of increasing the pendulum length because of the counterweight which can only move down. So, the system can relearn only those new external forces which have higher frequencies, not the lower ones. Even more, once the counterweight reaches the surface, the system stops functioning as a tunable resonator.

A possible way of preventing such a behavior is to allow $m$ to be a periodic function of time capable of repeating its tuning path as many times as needed. The simplest way is to consider $m$ as a function of angle. For example, representing $m$ as

$$m = a^2 + b^2 \cos^2 \theta \qquad (6.13)$$

will transform it into a periodic variable changing in the range between $a^2$ and $a^2 + b^2$. By appropriately selecting these parameters this range can be made as wide as we want. Making $m$ dependent on $\theta$ means promoting the latter to the role of the independent dynamical variable. Consider for example the Lagrange function of the form

$$L = \frac{(a^2 + b^2 \cos^2 \theta)\dot{x}^2}{2} - \frac{kx^2}{2} \tag{6.14}$$

In this case, the resonance condition will take the form

$$\cos \theta = \frac{\sqrt{k - \omega^2 a^2}}{\omega b} \tag{6.15}$$

and the equations of motion for the dynamic variable $\theta$ will read

$$\frac{b^2}{2} \sin 2\theta \, \dot{x}^2 = f_\theta - \gamma_\theta \dot{\theta} \tag{6.16}$$

This equation can similarly be averaged over time-window and, after using the notations () reduced to the optimal control form

$$\gamma_\theta \frac{\Delta \theta}{\Delta t} = f_\theta - \frac{1}{2\gamma} \langle \dot{x}^2 \sin 2\theta \rangle \tag{6.17}$$

In this case, if

$$f_\theta = \frac{1}{2\gamma} \max_\theta \langle \dot{x}^2 \sin 2\theta \rangle - \epsilon \tag{6.18}$$

with a small $\epsilon$, then we will have the same situation as described earlier. We will have a stable point approachable from below. As before this stable point can be disappear if $\epsilon$ suddenly drops below zero, however, in contrast with the previous case, in this case the arising instability will be temporary, and the system will tune-up to any external force with frequencies falling into the interval

$$\omega \in \left[ \frac{k}{\sqrt{a^2 + b^2}}, \frac{k}{a} \right] \tag{6.19}$$

In the absence of any external forces, the system may reside in the 'waiting mode' as long as needed, performing periodic motion

$$\gamma_\theta \frac{\Delta \theta}{\Delta t} = f_\theta \tag{6.20}$$

provided that the constant external force $f_\theta$ acting on the angular variable $\theta$ is present. This motion can be as slow as possible – this will only affect the learning and re-learning time.

In terms of the pendulum model we considered earlier, its angular modification can be realized as shown in the picture below:

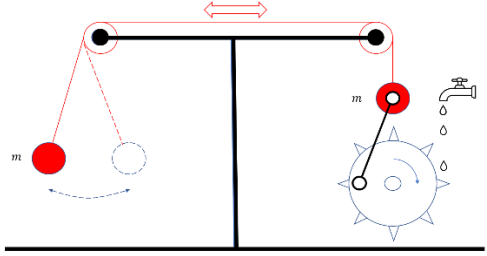

We see that in this case the masses of both pendulum and counterweight are made equal and the change of pendulum's length is achieved by applying an additional periodic force to the counterweight forcing the latter to slowly move up or down periodically changing the direction.

## 7. Self-Tunability: The N-Dimensional Case

In previous two sections we considered the memoryless case in which the learning outcomes have been achieved via self-tuning. Even these oversimplified and highly idealized examples we discussed above clearly showed that self-tuning is hard primarily because its long-lasting stability cannot be easily guaranteed. Any change in the external or internal condition may force the learning system to start re-learning an already learned material 'from scratch'. There is an obvious tradeoff between the quality of learning outcomes and their stability – the higher quality of learning can be achieved, the easier is to forget everything what has been learned. One can rephrase this statement in the following way: the harder is to learn something the easier is to forget it. In a sense, there is nothing bad about such statements because in many cases they reflect the reality: usually things are hard to learn if they are untypical, i.e., not widespread enough, so seeing them again is highly improbable. For such things the short-term memory would obviously suffice. However, learning without any memory at all is meaningless. So, we are naturally coming to the need for memory models. As noted earlier, for constructing such models one may need many resonators, which means that these models must be multi-dimensional.

Consider a $N$-dimensional harmonic oscillator described by the following Lagrangian

$$L = \frac{\dot{x}m\dot{x}}{2} - \frac{xkx}{2} \qquad (7.1)$$

in which $x$ and $\dot{x}$ are $N$-dimensional vectors and $m$ and $k$ are some symmetric and positive-definite $N \times N$ matrices. The equations of motion for this Lagrangian have the standard form:

$$m\ddot{x} + kx = f - \Gamma\dot{x} \qquad (7.2)$$

in which $\Gamma$ is a certain non-negative-definite $N \times N$ matrix describing possible dissipation in the system and $f$ is a vector of external forces. Introduce the new variables

$$y = Sm^{1/2}x \qquad (7.3)$$

with a certain orthogonal matrix $S$ diagonalizing the LHS of equation (7.2) and reducing it to the form:

$$\ddot{y} + \omega^2 y = \Gamma'(\phi - \dot{y}) \tag{7.4}$$

in which $\omega^2$ is a diagonal $N \times N$ matrix of system eigenfrequency squares, $\Gamma$ is another non-negative-definite matrix,

$$\Gamma' = S m^{-1/2} \Gamma m^{-1/2} S^{-1} \tag{7.5}$$

and $\phi$ is a vector related to the original vector of forces $f$ as

$$f = \Gamma m^{-1/2} S^{-1} \phi \tag{7.6}$$

The pure-resonant state would correspond to the situation when both the LHS and RHS of () are zero:

$$\ddot{y} + \omega^2 y = 0, \quad \phi = \dot{y} \tag{7.7}$$

This gives us the form of the external forces for which the system may have pure-resonant solutions:

$$\dot{y} = \phi = \{c_i \cos(\omega_i t + \delta_i)\}_{i=1}^{N} \tag{7.8}$$

This gives us the most general form of the original components of the force vector

$$f_i = \sum_{k,l,m} \Gamma_{ik} m_{kl}^{-1/2} S_{lm} c_m \cos(\omega_m t + \delta_m) \tag{7.9}$$

We see that all the components of the allowed forces are linear combinations of $N$ elementary harmonics. So, they are not completely independent. As a most interesting special case of this solution, consider the case when

$$\Gamma_{ik} = \gamma \delta_{i1} \delta_{k1} \tag{7.10}$$

This will give us

$$f_i = \delta_{iN} f, \quad f = \sum_{l,m} d_m \cos(\omega_m t + \delta_m), \quad d_l = \gamma \sum_m m_{Nl}^{-1/2} S_{lm} c_m \tag{7.11}$$

Introducing the notations: $x = x_N$, $\mu = m_{NN}$, $\mu_k = m_{Nk}$, $\kappa = k_{NN}$, $\kappa_k = k_{Nk}$, $k = 1, \ldots, N-1$, we arrive at the system of equations

$$\mu \ddot{x} + \kappa x + \sum_{k=1} \mu_k \ddot{x}_k + \sum_k \kappa_k x_k = f - \gamma \dot{x}, \tag{7.12}$$

$$\sum_{k=1}^{N-1} m_{ik} \ddot{x}_k + \sum_{k=1}^{N-1} k_{ik} x_k = 0, \quad i = 1, \ldots, N-1 \tag{7.13}$$

We see that only the first equation in the above system has the RHS, the RHSs of all other equations are zero. This is equivalent to saying that we want to consider only $x = x_N$ as the resonator variable, and keep other $N-1$ variables $x_1, \ldots, x_{N-1}$ are internal variables of the resonator. The pure resonant state can be achieved for any external force $f$ acting on the variable $x$ and having the form of any linear superposition of the $N$ basic harmonics of the resonator with frequencies $\omega_1, \ldots, \omega_N$. In other words, any such force can be learned by the resonator.

In case when the eigenfrequencies of (7.1) do not coincide with $\omega_b$, the question is if the system can tune itself to achieve such a coincidence. It is clear that this can be done via modification of matrix $m_{ik}$, and for that we need to associate it with dynamical variables. Since in this case we deal with matching two vectors, not two numbers, the number of angular variables should be at least equal to the dimension of that vector. Denote it by $N$. So, we consider $m_{ik}$ as a periodic function of $N$ new dynamical variables of angular type

$$m_{ik} = m_{ik}(\theta_1, \ldots, \theta_N) \tag{7.14}$$

which gives us the additional system of dynamical equations for them

$$\eta_a \dot{\theta}_a = -\phi_a + \frac{1}{2} \sum_{i,k} \frac{\partial m_{ik}(\theta_1, \ldots, \theta_N)}{\partial \theta_a} \dot{x}_i \dot{x}_k \tag{7.15}$$

The form of these equations is not too important to us. The most important thing is that whatever the dependence of $m_{ik}$ of these variables is, this function can always be linearized in the vicinity of their critical values $\theta_a^R$ at which the resonance occurs. We can write this as

$$\theta_a = \theta_a^R + \delta\theta_a \tag{7.16}$$

and

$$m_{ik}(\theta_1, \ldots, \theta_N) \approx m_{ik}^0 + m_{ik}^1 \delta\theta_1 + \cdots + m_{ik}^N \delta\theta_N \tag{7.17}$$

Substituting (7.17) into (7.15) we obtain

$$\eta_a \dot{\delta\theta}_a = -\phi_a + \frac{1}{2} \sum_{i,k} m_{ik}^a \dot{x}_i \dot{x}_k \tag{7.18}$$

Noting that the resonator variables $x_i$ implicitly depend on $\delta\theta_a$ and their amplitudes maximize when $\delta\theta_a = 0$, we can conclude that (7.18) represents a set of typical optimal control equations. The further reasonings will essentially repeat what was said in the previous section and lead to the equations

$$\eta_a \frac{\Delta\delta\theta_a}{\Delta t} = -\phi_a + \frac{1}{2} \left\langle \sum_{i,k} m_{ik}^a \dot{x}_i \dot{x}_k \right\rangle \tag{7.19}$$

which can be analyzed in the same way as the analogous one-dimensional equations (6.17). The analog of the condition (6.18) will have the form

$$\phi_a = \frac{1}{2} \left\langle \sum_{i,k} m_{ik}^a \dot{X}_i \dot{X}_k \right\rangle - \epsilon_a \tag{7.20}$$

where $X_i$ denote the resonant versions of functions $x_i$ and $\epsilon_a$ are small parameter whose sign coincides with the sign of the first term in the RHS of (7.20).

## 8. Towards the Learning Path Optimization Principle

In the previous two sections we described a method of building self-tunable multi-resonant systems and considered their simplest one- and multi-dimensional realizations. We showed that self-tunability can be achieved by extending the resonator with appropriately chosen controller capable of driving the natural evolution of the system towards semi-stable attractors lying near its resonant states and characterized near-the-maximum values of certain functionals bilinear in system velocities. The form of these functionals resembled that of the average dissipation terms which encouraged us to talk about the possible interpretability of the self-tunability in the language of maximal entropy production principle. However, strictly speaking, the examples we considered below do not confirm the equivalence of our constructions and the above principle (despite their obvious similarity). First of all, because our attractors are not stable – they are semi-stable and can easily be destroyed by sudden fluctuations in the value of the constant forces acting on controller's variables, and second, because there is no theorem stating that the only parameters to which the procedure of 'dynamization' could be applied are parameters defining system's kinetic term. Indeed, note that if, instead of dynamizing the mass $m$ in the Lagrangian (6.1) we would dynamize the stiffness $k$, the end result would be the same – we would again get a semi-stable attractor near the resonant state – but now it would be associated with the maximum of another functional, with the average of the potential term $\langle x^2 \rangle$, not with the average of the kinetic term $\langle \dot{x}^2 \rangle$, as we had in section 6. This suggests that associating learning process exclusively with the maximal entropy production principle (i.e., with maximum dissipation rate) would not be quite correct, because, as we just noted, there could be many other functionals having different form but also achieving their maxima near the resonant state.

In this connection it is natural to ask ourselves how general the trajectory optimization methods similar to what we described in sections 6 and 7 could be, and how far they could extend beyond the simplest systems, resonator-controller pairs and forces we considered there. Are there any other methods which, by their very design, could cover a broader set of possible self-learning systems placed in a wider range of unspecified environments. Here we are entering a grey area of speculations. Nevertheless, we want to share some ideas, currently having status of unconfirmed hypotheses, which, as we think, could still give us some clue on possible ways of answering the above question. The main idea is to choose an appropriate formal language compatible with both the languages of ML and theoretical physics and allowing us to study the autonomous intelligent systems and abstract physical models within the same mathematical framework.

Let us consider a system of interacting point-like mechanical particles. As usually, we can characterize these particles by their positions $x$ and velocities $\dot{x}$ and represent the total energy of the system as a certain Hamiltonian function of $x$, $\dot{x}$ and $t$:

$$E = H(x, \dot{x}, t) \qquad (8.1)$$

We included the explicit dependence on time in (8.1) as the system can by assumption be open and its energy is not necessarily a conserved quantity. To take into account this non-conservativity on the microscopical (i.e., on an individual particle) level, assume that the particle can change its energy in some additional non-mechanical ways, say, by losing it through dissipation or gaining it via some external channels not directly describable within the Lagrange or Hamilton pictures. It is convenient to represent these non-mechanical energy changes as inward and outward energy flows entering and leaving the particle, respectively, as shown in the picture below.

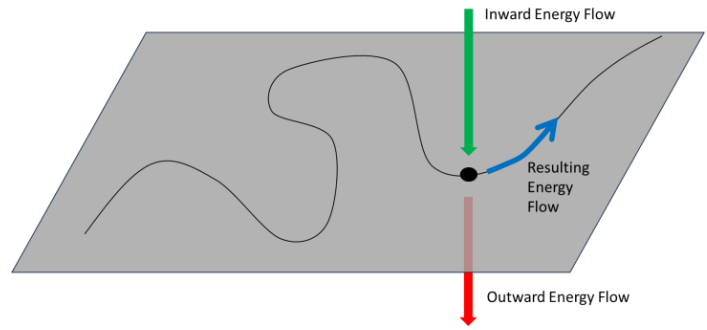

The idea is to consider the resulting energy flow (defined as the difference between the inward and outward flows) as an extra energy accumulated by the mechanical particle in a unit of time. Let us denote this resulting flow of energy, which can be considered as a certain function of its position $x$, velocity $\dot{x}$, and time $t$, by

$$\Lambda = \Lambda(x, \dot{x}, t) \tag{8.2}$$

Now note that the total energy accumulated by this particle while moving along some trajectory $x = x(t)$ can be defined by the time-integral of the above function. But from the energy conservation law it follows that this integral should be equal to the mechanical energy given by formula (8.1). It is worth stressing here that although we allow for a particle to gain or lose the energy in non-mechanical ways, we still want to consider the particle as a mechanical object. This means that once its energy changes in whatever ways, this change should immediately be reflected in particle's kinetic and potential energies constituting the function (8.1). This leads us to the equation

$$\frac{dH(x, \dot{x}, t)}{dt} = \Lambda(x, \dot{x}, t) \tag{8.3}$$

The form of the above equation is covariant under the following two simultaneous transforms:

$$H \to H + \Phi, \qquad \Lambda \to \Lambda + \frac{d\Phi}{dt} \tag{8.4}$$

in which $\Phi = \Phi(x, t)$ is an arbitrary function. This covariance is a consequence of the fact that the Hamiltonian of an open system is not defined uniquely – only up to the interaction terms (the interface), which can be attributed to both the system, its environment or both. This split can, obviously, be done in infinitely many ways, and this freedom is exactly what is reflected in equations (8.4).

Now assume that we want to find trajectories $x = x(t)$ maximizing a certain functional

$$D = \int_{t_0}^{t} \Delta(x, \dot{x}, \tau) d\tau \tag{8.5}$$

having the meaning of energy and corresponding physical dimension. In this case, the dimension of the sub-integral function $\Delta$ is [energy/time]. Although we denoted this functional by $D$ – the letter previously used for denoting the dissipated energy – its concrete meaning can be any, including the energy accumulated by the time $t$ (in the latter case $\Delta = \Lambda$). To correctly state the variational problem, we need to interpret the condition (8.3) as a scalar constraint on the possible trajectories of the system. If system is one-dimensional, i.e., the number of particles is equal to the number of conditions (in this case 1), then the constraint does not leave any freedom and we get the true equations of motion. However, for $N$-dimensional systems the number of the remaining degrees of freedom is $N - 1$, and these degrees can be subject to some additional macroscopic variational principles. In that case we can use the Lagrange multiplier, which, in this case must be some function of $t$. Denoting it by $\lambda = \lambda(t)$ we can write the extended Lagrange function with $N + 1$ dynamical variables (one variable $\lambda$ and $N$ variables $x$ as):

$$R(x, \dot{x}, \lambda, \dot{\lambda}, t) = \Delta(x, \dot{x}, t) + \lambda \Lambda(x, \dot{x}, t) + \dot{\lambda} H(x, \dot{x}, t) \tag{8.6}$$

The extremum (maximum) of this extended Lagrangian can be found in standard variational ways

$$\delta \left( \int_{t_0}^{t} R(x, \dot{x}, \lambda, \dot{\lambda}, \tau) d\tau \right) = 0, \quad x|_{t_0} = x_0 \tag{8.7}$$

Note that the variation by $\delta \lambda$ will give us the constraint (8.3) as a single equation, while the variation by $\delta x$ will lead to the system of $N$ additional second-order equations of motion. As a result, we get a system of Newtonian equations describing by construction the resource optimization process. It is important that the above variational principle leaves the upper integration limit free of any constraints. This makes the above principle local, in contrast to the principle of the least action, which is a global principle optimizing the entire trajectory and requiring specification of both its initial and final points. The locality of the above principle lies in the fact that at any instance of time $t$ the particle has a freedom to decide where to move next and selects the direction in which the increase of the functional $D$ is maximal.

To illustrate the above schema, let us consider a point-wise intelligent agent living on a certain surface characterized by a coordinate $x$. We can think of this surface as of a certain oil-rich terrain. The amount of energy (fuel) the agent can suck from the point $x$ during the infinitesimally small time between $t$ and $t + dt$ is given by $dE = U(x, t)dt$. Assume that the agent can move in that terrain. The motion (as any other change – in this case the change of position) cannot be done without spending some energy. The amount of energy $dE$ needed to move for an infinitesimally small distance from $x$ to $x + dx$ must be proportional to agent's velocity $\dot{x} = |dx|/dt$ and the distance itself $|dx|$. This will give us $dE = -(\mu/2)\dot{x}^2 dt$. The coefficient of proportionality in this formula has the meaning of the friction coefficient $\gamma = \mu/2$). Combining these two (positive and negative) changes of agent's internal energy we obtain the energy accumulation rate:

$$\Lambda(x, \dot{x}, t) = U(x, t) - \frac{\mu \dot{x}^2}{2} \tag{8.8}$$

This expression can be interpreted in two different ways. One way would be based on the that an intelligent agent is interested in accumulating as much of energy as possible. In that case, it would be looking for those trajectories $x = x(t)$ along which the total accumulated energy, i.e., a time integral of function (8.8)

$$E = \int_{t_0}^{t} \left( U(x, t) - \frac{\mu \dot{x}^2}{2} \right) d\tau \tag{8.9}$$

would be maximal. This suggests that the problem of finding such trajectories can be solved by using the variational method and would lead to the standard problem similar to that of minimizing the action functional. This idea is supported by the observation that function $\Lambda$ looks like the difference between the potential and kinetic energies, so its form resembles the form of the Lagrange function taken with sign minus. Therefore, the fact that is must be maximized, rather than minimized, looks quite natural. Another circumstance favoring the idea of applying the variational principle to the integral of function $\Lambda$ is that it is not a Lagrange function (even if considered with the reverted sign) because the two terms in (8.9) represent the energy flows, not the energies. As to the integral itself, now it has the meaning of the energy, not the action, as it would be in the case of mechanics. This solves the problem we discussed in the beginning of this section – the problem of 'incorrect' dimension of the 'resource' that is being optimized in the principle of the least action. Now the resource we are trying to optimize has correct dimension of energy, and thus is fully compatible with the energetic principles of learning. Does it mean that we are ready to state the resource optimization problem for our intelligent agent in the form

$$\delta \int_{t_0}^{t} \left( U(x,t) - \frac{\mu \dot{x}^2}{2} \right) d\tau = 0 \ ? \tag{8.10}$$

The answer is 'not yet', and here is why. The point is that the agent is not completely free and cannot absorb as much energy as the variation problem (8.10) would allow. The point is that agent's energy storage capacity is limited, because the agent is by assumption a point-wise mechanical object, and therefore the only way for it to store the acquired energy is to somehow distribute it between its own kinetic and potential energies. In other words, agent's energy should have the purely mechanical form:

$$H(x, \dot{x}, t) = V(x,t) + \frac{m\dot{x}^2}{2} \tag{8.11}$$

where $V(x,t)$ is agent's potential energy and $m$ is its mass. In this case the general relation (8.3) takes the form

$$V(x,t) + \frac{m\dot{x}^2}{2} = E = \int_{t_0}^{t} \left( U(x,\tau) - \frac{\mu \dot{x}^2}{2} \right) d\tau \tag{8.12}$$

and plays the role of a constraint on the possible forms of agent trajectories $x(t)$. We can also rewrite the integral relation (8.12) in the differential form as

$$\frac{d}{dt}\left( V(x,t) + \frac{m\dot{x}^2}{2} \right) = U(x,t) - \frac{\mu \dot{x}^2}{2} \tag{8.13}$$

It is easily seen that in a particular case when

$$V(x,t) = W(x) - f(t)x, \quad U(x,t) = -\dot{f}(t)x \tag{8.14}$$

where $f(t)$ is a certain function of time, $\gamma = \mu/2$, and energy is defined as a function of coordinates and velocities only

$$E = V(x) + \frac{m\dot{x}^2}{2} \tag{8.15}$$

with no explicit dependence on time, we will arrive at the old relation

$$\dot{E} = f\dot{x} - \gamma\dot{x}^2 \tag{8.16}$$

we discussed earlier in section 3 (see formula (5)).

## 9. Discussion. Why Classical Mechanics?

In this article, we tried to consider the nature of learning – a key component of intelligence – from the point of view of physics. As a physical basis – a potential platform for discussing the physical aspects of the learning process – we used the formalism of classical mechanics. Why? Because the language of classical mechanics can be seen as a natural bridge between the languages of cognitive science, machine learning and physics. There are several reasons for this.

1. The most common reason, not directly related to machine learning or cognitive science, is based on the fact that classical mechanics lies in the foundation of practically all physics. This means that any natural phenomenon allowing even an approximate or qualitative explanation in the language of classical mechanics, has a big chance to be fully and quantitatively explained within one of the branches of mathematical/theoretical physics.

2. Classical mechanic actively exploits the concept of the resource naturally arising within its formalism in the form of energy. The global energy conservation law makes the evolution of any open mechanical system similar to a zero-sum game between that system and its environment (or other systems). This implicitly endows any open mechanical system with features usually attributed to diverse life-forms because the need for resource is the central thing that characterizes life and makes it purposeful. This (main) purpose then cascades down and transforms into myriads of different particular forms of intelligent behavior. The phrases like: 'the system performs a certain action to accumulate resource', or 'system has accumulated enough resource to perform some actions' – can be addressed equally well to both intelligent agents (like humans) and non-living mechanical systems blindly following equations of classical mechanics. The fact that the ability of acting is conditioned upon the availability of resource, makes the search for resource the most meaningful action for any resource-driven autonomous system. This explains the cyclicity of dynamics of both mechanical systems (like pendulums) and living organisms (in their daily routine) and establishes strong parallels between their behavior.

3. Classical mechanics (at least, in its Newtonian form) is capable of simulating virtually any computation no matter how complex is it. To illustrate this statement, remember that computation, including its most sophisticated forms such as the ML and AI algorithms, can be viewed as the evolution of dynamical system. The role of the current state of the system is played by a certain numerical vector $z(t)$ describing the content of all its memory registers at time $t$. The next state $z(t + \delta t)$ is determined by the collection of rules (constituting the algorithm or program) showing how to update the current content. Here $\delta t$ is the elementary update time. We can always represent the state updating rule in the vector form: $z(t + \delta t) = z(t) + \delta t \cdot r(z(t), t)$, where $r(...)$ denotes a vector rule-function – the instruction set, in other words. We added explicit time-dependence to it because the instruction set may be dependent on the external conditions – user input or any other changes in external data. In today's computers, both $\delta t$ and $z(t)$ are discrete quantities, however, in case when $\delta t$ is small and the measure of $z$'s discontinuity is small too, the above discrete equation can be approximated by its continuous version $\dot{z}(t) = r(z(t), t)$, which represents the system of continuous first-order dynamical equations of the most general form. By differentiating both sides of this system by $t$ we can reduce it to the second-order form, $\ddot{z}(t) = q(z(t), \dot{z}(t), t)$, in countless number of ways, each of which will gives us a certain system of Newtonian equations in the most general form.

4. Another interesting feature, typical for all the intelligent living organisms and implicitly present in the mathematical formalism of classical mechanics (at least in its Lagrange version), is the ability to optimize the consumption and spending of system resources. We mean here the 'principle of least action', according to which the mechanical system always choses those trajectories of evolution along which a special resource, usually referred to as 'action' and defined as the time integral of the difference between the kinetic and potential energies, is minimal. This principle endows a mechanical system with a sort of 'laziness' – another human-like feature expressed as the avoidance of resource-consuming actions and preference of spending more time in resource-rich areas. Also, this principle enforces the mechanical system to do some 'planning' before moving – as the optimization problem can be correctly stated and solved only after specifying in advance where the system wants to be at certain future time. Although attributing to mechanical systems the ability to plan and optimize is no more than a convenient language for describing some aspects of their behavior, we tend to consider any physical theory that allows such language as a potentially interesting candidate for the role of a future theory capable of describing the phenomenon of intelligence.

Points 3 and 4 seem to us the most interesting and important, but there is one problem. The fact is that the Newtonian and Lagrangian versions, in which these points manifest themselves in the most natural ways, are equivalent only in the case of closed and stationary systems. If the system is open, this equivalence is violated. Since we are primarily interested in open systems, it is very important for us to fully understand the natural boundaries of each of these versions. Both have their pros and cons.

Indeed, despite the obvious appeal of Lagrange's formalism, which allows us to treat mechanics as a problem of optimizing resources within the framework of the principle of least action, this principle cannot be directly and in the same way used to describe learning processes. One reason for this is that the action - the quantity that plays the role of an optimized resource in the above principle - is not the same resource that should be optimized in the case of learning-related tasks. Indeed, the physical dimension of action is 'energy' × 'time', while, as we have seen in the previous sections, the optimized resource for learning tasks must be energy itself.  Another reason is that Lagrange's version, at least in its standard form, cannot describe dissipation, which, as we have shown above, plays a crucial role in explaining the phenomenon of learning. From this point of view, Newton's version of classical mechanics seems to be the best choice. Indeed, as we have seen above, the latter is directly related to computation (and through it to all machine learning algorithms), and its equations can easily be used to describe essentially non-conservative systems. At first glance, this is exactly what we need. However, Newtonian mechanics is 'too formal' and (unlike Lagrange's formalism) does not speak the 'language of goals'. Since the concept of optimization is central to all learning algorithms and has been directly or indirectly present in all of our previous discussions, it would be a great pity that it cannot be used naturally (i.e., as an organic component of the mathematical formalism used to describe the physical mechanisms of learning).

This brings us to the idea of combining Newton's and Lagrange's formalisms into a single formalism. In sections 4-7 we have already tried to use some elements of the Lagrange formalism in the Newtonian schema, explicitly separating the conservative and dissipative terms and placing them on the left and right sides of the equations of motion, respectively. However, this was not done to increase the interpretability of dynamic equations from the point of view of optimization problems, but simply to compactify some formulas and their conclusions.

A more meaningful and self-consistent method of such unification was proposed in [20], where we showed that any first-order dynamical system, regardless of whether it is conservative or not, can be formally described within the framework of the standard Lagrange formalism based on Lagrange functions that are linear over the velocities of the system. Due to the mutual convertibility of first-order dynamical equations into second-order Newtonian equations and vice versa, this method actually made it possible to consider Newtonian mechanics within the framework of the Lagrange formalism. However, the linearity of the Lagrange function in the velocities of the system made it difficult to associate any maximizable functionals with it.

The approach presented in Section 8 has a good chance of being free of the aforementioned problem, which gives us some hope that it may eventually develop into another, conceptually purer and more universal way of combining these two formalisms into one.

And a few words in conclusion. We often say that information has value, and we often intuitively treat it as a resource. It turns out that the validity of this statement goes far beyond the standard examples from everyday life: it has much deeper origins, lying in the most fundamental laws of nature, usually studied by various branches of theoretical physics. There is a direct correspondence between the information as we know it from Shannon's theory and the energy as we know from physics. It can be proved that any correctly predicted phenomenon becomes an energy resource. This simple fact, which was explained and discussed in detail in [17], not only immediately makes it clear on an intuitive level why even primitive living organisms may 'want' to learn, but also tells us about the coexistence of two different but essentially equivalent scientific languages for solving the AI problems: the language of information theory and the language of physics (in the case of [17] it was thermodynamics). These two languages complement and enrich each other, and the very idea of combining them into a single and more powerful metalanguage can have enormous potential for a better understanding of the phenomenon of intelligence.

Interestingly, in this article, we essentially came to the same conclusion, but from a completely different physical point of view. We considered the problem of learning within the framework of the formalism of classical mechanics, which, as we know, does not allow the use of the concept of information in the standard (i.e., probabilistic, or Shannonian) sense, simply because classical mechanics is a purely deterministic theory. However, we can use the term 'information' informally, implying by it a measure of the representability of the observed data as a combination of some already known basic patterns (for example, the elementary Fourier harmonics). Having done this, we immediately arrive at a similar relationship between information and energy as we had in case of thermodynamics: the more regular the external force, the more energy can be extracted from it by means of the resonance effect. Observation of this similarity was the main reason why we decided to explore the possibility of using the mathematical basis of classical mechanics to tie together the problems of physics and AI. Fully realizing that it would be too naïve to literally consider classical mechanics as a direct candidate for the role of the definitive unified metalanguage, we still hope that some of the ideas proposed in this article can eventually develop into a more valid theory of intelligence and thus pave the way for practical advances in this field.

# References

[1] Marion Blue, "Learning theory and the evolutionary analogy", Erindale Coll., Uni. of Toronto (1979)
[2] Vitaly Vanchurin, Mikhail I. Katsnelson and Eugene V. Koonin, "Toward a theory of evolution as multilevel learning", Proc. Natl. Acad. Sci. U.S.A. 119, 10.1073/pnas.2120037119 (2022)
[3] P. Villalobos and A. Ho, "Trends in training dataset sizes," https://epochai.org/blog/trends-in-trainingdataset-sizes, 2022, accessed: 2022-09-27.
[4] J. Kaplan et al., "Scaling laws for neural language models,", arXiv:2001.08361, 2020.
[5] Ian Goodfellow, Yoshua Bengio, and Aaron Courville, "Deep Learning", MIT Press, https://www.deeplearningbook.org/, (2016).
[6] Neil C. Thompson, Kristjan Greenewald, Keeheon Lee, Gabriel F. Manso, "The computational limits of deep learning", MIT Initiative on the digital economy research, https//ide.mit.edu/wp-content/uploads/2020/09/RBN.Thompson.pdf, (2020)
[7] Fabio Pardo, Arash Tavakoli, Vitaly Levdik, Petar Kormushev, "Time Limits in Reinforcement Learning", arXiv:1712.00378 (2018)
[8] Zhongkai Hao, Songming Liu, Yichi Zhang, Chengyang Ying, Yao Feng, Hang Su, Jun Zhu, "Physics-Informed Machine Learning: A Survey on Problems, Methods and Applications", arXiv:2211.08064v2 [cs.LG] 2023
[9] G. E. Karniadakis, I. G. Kevrekidis, L. Lu, P. Perdikaris, S. Wang, and L. Yang, "Physics-informed machine learning," Nature Reviews Physics, vol. 3, no. 6, pp. 422–440, 2021.
[10] N. Thuerey, P. Holl, M. Mueller, P. Schnell, F. Trost, and K. Um, "Physics-based deep learning," arXiv preprint arXiv:2109.05237, 2021.
[11] C. Beck, M. Hutzenthaler, A. Jentzen, and B. Kuckuck, "An overview on deep learning-based approximation methods for partial differential equations," arXiv preprint arXiv:2012.12348, 2020.
[12] S. Cuomo, V. S. Di Cola, F. Giampaolo, G. Rozza, M. Raissi, and F. Piccialli, "Scientific machine learning through physicsinformed neural networks: Where we are and what's next," arXiv preprint arXiv:2201.05624, 2022.
[13] S. Cai, Z. Mao, Z. Wang, M. Yin, and G. E. Karniadakis, "Physicsinformed neural networks (pinns) for fluid mechanics: A review," arXiv preprint arXiv:2105.09506, 2021.
[14] M. Lutter, J. Silberbauer, J. Watson, and J. Peters, "Differentiable physics models for real-world offline model-based reinforcement learning," in 2021 IEEE International Conference on Robotics and Automation (ICRA). IEEE, 2021, pp. 4163–4170.
[15] C. Xie, S. Patil, T. Moldovan, S. Levine, and P. Abbeel, "Modelbased reinforcement learning with parametrized physical models and optimism-driven exploration," in 2016 IEEE international conference on robotics and automation (ICRA). IEEE, 2016, pp. 504– 511.
[16] Barve, Aditya; Wagner, Andreas, "A latent capacity for evolutionary innovation through exaptation in metabolic systems" . Nature. 500 (7461): 203–206, 2013

[17] Alex Ushveridze, "Can Turing machine be curious about its Turing test results? Three informal lectures on physics of intelligence", arXiv:1606.08109, (2016)
[18] Carlos Gershenson, Vito Trianni, Justin Werfel, Hiroki Sayama, "Self-Organization and Artificial Life", arXiv:1903.07456 [nlin.AO], (2020)
[19] Martyushev, L.M.; Seleznev, V.D. Maximum entropy production principle in physics, chemistry and biology. Phys. Rep. 2006, 426, 1–45.
[20] Alex Ushveridze, "Classical Lagrange formalism for non-conservative dynamical systems", arXiv:2212.12409, (2022)

## Appendix. Pure Resonant States in One-Dimensional Systems

In this Appendix we will try to describe a maximally general class of external forces $f(t)$ for which a one-dimensional Lagrangian $L$ would allow a pure resonant solution nullifying both RHS and LHS of the dynamical equation

$$\frac{d}{dt}\frac{\partial L}{\partial \dot{x}} - \frac{\partial L}{\partial x} = f - \gamma \dot{x} \qquad (A.1)$$

Let us start with an obvious remark that if the RHS of the equation (A.1) is identically zero, then the equation (A.1) becomes indistinguishable from the equation of motion for a certain closed and conservative system described by Lagrangian $L$. The conservativity means two important things: 1) that the energy of the system is a conserved quantity and 2) that the system allows a fully equivalent description in terms of the energy $E(x, v)$ defined by the standard formula (4.2) and considered as a function of coordinate and velocity.

The equivalence between the energy-based and Lagrange pictures allows us to uncover the $f \to L$ relationship in two stages: first solving an easier problem of establishing the $f \to E$ relationship, and then, after the energy function is known, reconstruct the Lagrange function $L$ by simply inverting the relation (4.2). The corresponding inversion formula reads:

$$L(x, \dot{x}) = v \int \frac{E(x,v)}{v^2} dv \bigg|_{v=\dot{x}} \qquad (A.2)$$

As the departure point, we will use the fact that according to (2.5) the velocity is proportional to the external force $f(t)$ which is assumed to be a given function of time $t$.

Let us consider a certain time interval $[t_{min}, t_{max}]$ and define in it a continuous function $f(t)$ oscillating around its zero value, $f = 0$, in some (possibly irregular) ways. Let $t_n$, $n = 0, \ldots, N$ denote the zeros of this function in that interval:

$$f(t_n) = 0, \quad n = 0, 1, \ldots, N; \qquad t_0 = t_{min}, \quad t_N = t_{max} \qquad (A.3)$$

and let $f_n(t)$ denote distinct sign-definite parts of function $f$ in the intervals $[t_{n-1}, t_n]$. The oscillatory character of $f$ means that the signs its neighboring parts $f_{n-1}(t)$ and $f_n(t)$ should be opposite. The allowed irregularity of these oscillations implies that the internal shape of these parts could be any. We will impose only one limitation on these shapes requiring that the areas under the curve of function $f(t)$ between its zero point $t_n$ are exactly the same:

$$\int_{t_{n-1}}^{t_n} f_n(t)dt = (-1)^n A, \quad n = 1, \ldots, N \tag{A.4}$$

We allow the function $f(t)$ to extend beyond the interval $[t_{min}, t_{max}]$ in either direction by combining its parts $f_n$ defined in $[t_{min}, t_{max}]$ into new sequences. We can think of $f_n$ as characters forming a certain alphabet. Then the sequences of these parts can be thought as messages encoded through a simple change of the order of these characters. This means that we are not requiring the preservation of the original order the parts $f_n$ they had in the 'mother' interval $[t_{min}, t_{max}]$. Outside of this interval, it can be any, provided that the sign-alternating character of the sequence is preserved.

From the fact that the sign of function $f(t)$ does not change between its zeros it follows that $x(t)$ should be a monotonic function of $t$ in each of the $N$ intervals $[t_{n-1}, t_n]$. This means that when $t$ changes between $t_{n-1}$ and $t_n$ the coordinate $x$ monotonically changes in some interval $[x_{n-1}, x_n]$. For the endpoints of this interval, we may have either $x_{n-1} < x_n$ or $x_{n-1} > x_n$, depending on the increasing or decreasing character of this monotonicity. These endpoints are obviously the turning points at which the monotonic motion in one direction switches to the monotonic motion in the opposite direction.

Now, we know that any function $x(t)$ monotonous in a certain interval must be invertible in that interval. Since we have $N$ different intervals this gives us $N$ different inverse functions

$$t = t_n(x) \in [t_{n-1}, t_n], \quad x \in [x_{n-1}, x_n], \quad n = 1, \ldots, N \tag{A.5}$$

Note that no unique inverse function of $x(t)$ can exist because $x(t)$ is not a globally monotonic function, it is only sequentially monotonic. However we can informally talk of (A.5) as of $N$ branches of a single but multivalued function $t(x)$. An immediate consequence of a multi-valued character of $t(x)$ is that the velocity if considered as a function of coordinates $x$ also becomes a multi-valued function whose multiple branches are given by

$$v_n(x) = \gamma^{-1} f_n(t_n(x)) \tag{A.6}$$

Substituting these expression into the energy conservation condition we can write

$$E(x,v) = E = const \tag{A.7}$$

We need to find such a form of $E(x,v)$ for which (A.7) would automatically be satisfied for all the branches $v = v_n(x)$. In other words, the LHS of (A.7) must be independent on both $x$ and $n$ simultaneously. Fortunately, finding such expressions is easy and there are many possibilities to do that. The simplest one is based on the expression

$$E(x,v) = E + v^M \prod_{n=1}^{N}(v - v_n(x)) \tag{A.8}$$

This expression can equivalently be rewritten in the form

$$E(x,v) = E + v^M \sum_{n=0}^{N} \sigma_n(x) v^{N-n} \tag{A.9}$$

where

$$\sigma_0(x) = 1, \quad \sigma_1(x) = \sum_{n=1}^{N} v_n(x), \quad \ldots, \quad \sigma_N(x) = \prod_{n=1}^{N} v_n(x) \qquad (A.10)$$

denote the elementary symmetric polynomial of order $n$ of functions $v_1(x), \ldots, v_N(x)$.

From the equations (A.9) and (A.10) it follows that $N + M$ must be even because the energy represented by odd-degree polynomial in velocity $v$ cannot be bounded from below. The first obvious limitation for the form of function $f(t)$ also comes from the expression (A.9). To be well defined on a certain support interval $[x_{min}, x_{max}]$ all the velocity branches $v_n(x)$ must have the same support interval. This is possible only if the areas under the curve of function $f(t)$ between its zero point $t_n$ are exactly the same. This is the first limitation we are forced to impose on function $f(t)$. Substituting the expression for energy into formula (A.2), and performing trivial integration over $v$ we obtain the Lagrangian of the system

$$L(x, \dot{x}) = \sum_{n=0}^{N} \sigma_n(x) \dot{x}^{N+M-n} / (N + M - n - 1) \qquad (A.11)$$

This expression is well defined if $M \geq 2$, because of the guaranteed absence of the potentially dangerous division by 0. However, we can safely consider cases with $M = 0$, and not worry too much about that if this division occurs at the term linear in velocity, because in this case it is a total time-derivative and thus should not have any effect on the equations of motion. However, if we still want to clean-up this case, we can do that by imposing an additional symmetry condition on function $f(x)$ which would automatically ensure that $\sigma_{N-1}(x) = 0$ but not affect the other coefficients $\sigma_n(x)$ with $n \neq N - 1$. This can be done if one uses the relation:

$$\sigma_{2N-1}(x) = \left( \frac{1}{v_1(x)} + \cdots + \frac{1}{v_{2N}(x)} \right) \sigma_{2N}(x) = 0 \qquad (A.12)$$

and requires that the sum of the velocity inverses is 0. This is equivalent to requiring

$$\frac{1}{f(t_1(x))} + \cdots + \frac{1}{f(t_N(x))} = 0 \qquad (A.13)$$

So, the Lagrange function for the external force is constructed. This completes the proof.

Consider as an example the case with $M = 0$ and $N = 2$ with strictly periodic function $f(x)$ each period of which has a form of two waves satisfying the following conditions

$$f(t_1 - t) + f(t_1 + t) = 0 \qquad (A.14)$$

and thus corresponding to the case with two branches, $v_1(x)$ and $v_2(x)$. It is easily seen that because of (A.14) the condition (A.12) is automatically satisfied

$$\frac{1}{f(t_1(x))} + \frac{1}{f(t_2(x))} = 0 \qquad (A.15)$$

and, in addition to that, we have

$$v_2(x) = -v_1(x) \tag{A.16}$$

which brings us back to the case we considered in section 5 in connection with the the periodic anharmonic case.